\documentclass{emulateapj}

\def\gtaprx {\lower .1ex\hbox{\rlap{\raise .6ex\hbox{\hskip .3ex
	{\ifmmode{\scriptscriptstyle >}\else
		{$\scriptscriptstyle >$}\fi}}}
	\kern -.4ex{\ifmmode{\scriptscriptstyle \sim}\else
		{$\scriptscriptstyle\sim$}\fi}}}
\def\ltaprx {\lower .1ex\hbox{\rlap{\raise .6ex\hbox{\hskip .3ex
	{\ifmmode{\scriptscriptstyle <}\else
		{$\scriptscriptstyle <$}\fi}}}
	\kern -.4ex{\ifmmode{\scriptscriptstyle \sim}\else
		{$\scriptscriptstyle\sim$}\fi}}}

\newcommand{\cutt}[1]{\textcolor{blue}{}}

\newcommand{\Ni}{{\ensuremath{^{56}\mathrm{Ni}}}}

\newcommand{\Co}{{\ensuremath{^{56}\mathrm{Co}}}}

\begin{document}

\title{The Los Alamos Supernova Light-Curve Project: Computational Methods}

\author{Lucille H. Frey\altaffilmark{1,2}, Wesley Even\altaffilmark{2}, 
Daniel J. Whalen\altaffilmark{3}, Chris L. Fryer\altaffilmark{4,5,6}, Aimee 
L. Hungerford\altaffilmark{2}, Christopher J. Fontes\altaffilmark{7}, and James Colgan\altaffilmark{8}}

\altaffiltext{1}{Department of Computer Science, University of New Mexico,
Albuquerque, NM  87131, USA}

\altaffiltext{2}{XTD-6, Los Alamos National Laboratory, Los Alamos, NM 87545, USA}

\altaffiltext{3}{Department of Physics, Carnegie Mellon 
University, Pittsburgh, PA 15213, USA}

\altaffiltext{4}{CCS-2, Los Alamos National Laboratory, Los Alamos, NM 87545, USA}

\altaffiltext{5}{Physics Department, University of Arizona, Tucson, AZ 85721, USA}

\altaffiltext{6}{Physics and Astronomy Department, University of New Mexico, Albuquerque, NM 87131, USA}

\altaffiltext{7}{XCP-5, Los Alamos National Laboratory, Los Alamos, NM 87545, USA}

\altaffiltext{8}{T-1, Los Alamos National Laboratory, Los Alamos, NM 87545, USA}

\begin{abstract}

We have entered the era of explosive transient astronomy, in which current and upcoming real-time surveys such as the 
Large Synoptic Survey Telescope (\textit{LSST}), the Palomar Transient Factory (\textit{PTF}) 
and Panoramic Survey Telescope and Rapid Response System (\textit{Pan-STARRS}) will detect supernovae in 
unprecedented numbers.  Future telescopes such as the James Webb Space Telescope may discover supernovae from 
the earliest stars in the universe and reveal their masses.  The observational signatures of these 
astrophysical transients are the key to unveiling their central engines, the environments in which they occur, and 
to what precision they will pinpoint cosmic acceleration and the nature of dark energy.  We present a new method 
for modeling supernova light curves and spectra with the radiation hydrodynamics code RAGE coupled with detailed 
monochromatic opacities in the SPECTRUM code.  We include a suite of tests that demonstrate how the improved 
physics and opacities are indispensable to modeling shock breakout and light curves when radiation and matter are 
tightly coupled.  

\end{abstract}

\keywords{atomic data - methods: numerical - radiative transfer - supernovae: general}

\maketitle

\section{Introduction} \label{sec:intro}
We are now in the era of explosive transient astronomy, in which real-time all-sky surveys with the 
Large Synoptic Survey Telescope \citep{LSST}, the Palomar Transient Factory 
\citep{PTF} and Panoramic Survey Telescope and Rapid Response System 
\citep{PSTARRS} are or soon will be detecting supernovae (SNe) in unprecedented numbers.  
Gamma-ray bursts 
now probe the universe out to $z \gtrsim 8$ and upcoming telescopes such as the James Webb Space Telescope 
\citep{JWST} could discover SNe from the 
earliest stars (Pop III SNe) and reveal their masses.  The observational signatures of these cosmic 
explosions are key to unveiling their central engines, the environments in which they occur, and to what precision 
they can be used as standard candles to pinpoint cosmic expansion rates and constrain the nature of dark energy.  
Light curves only provide indirect measures of the physics at work in transients but are by far the most abundant 
type of data.  New advances in numerical codes can now produce SN light curves and spectra with greater accuracy and 
time resolution than in previous studies, mandating a new generation of simulations to better understand these cosmic 
explosions.

Numerous codes have been developed in the past three decades to simulate SNe and their light curves and
spectra, with varying degrees of physics.  They include codes with radiation transport linked to simple hydrodynamics 
such as EDDINGTON \citep{EP93} and others \citep{Hoeflich93}, one-temperature radiation diffusion codes such as 
Kepler \citep{Weaver1978,Woosley2002}, and radiation hydrodynamics (RHD) codes that evolve radiation and material 
temperatures separately such as VISPHOT \citep{Ens92}, TITAN \citep{Gehmeyr94} and STELLA \citep{Blinn98}.  
There are many codes which have been developed to post-process profiles from hydrodynamical codes with radiation transport, 
including SEDONA \citep{Kasen09}, SYNOW, \citep{Branch85} and several Monte Carlo codes (\citealt{ML93}, \citealt{KSim09} and others).  
RHD simulations have also been post-processed to obtain non-LTE (local thermodynamic equilibrium) spectra \citep{HE94}.

In this paper we present a new method developed at Los Alamos National Laboratory (LANL) in which we start from a full 
RHD simulation and post-process the resulting structure with detailed monochromatic opacities to 
calculate SN light curves and spectra 
with an unprecedented level of accuracy.  This level of detail is necessary for times when radiation-matter
coupling is important, from shock breakout until the start of the nebular phase when the SN ejecta become optically thin.  
Models with full RHD are required where radiation plays an important role in the hydrodynamics, 
such as where shock heating is important in determining the temperature profile.  The codes presented here do not incorporate 
non-LTE effects, which become more significant at late times in SNe.  

In the initial stages of supernovae, rebound from gravitational collapse or thermonuclear burning drives a highly 
radiative shock into the upper layers of the star. The shock is not visible until it 
breaks through the surface of the star because opacity due to electron scattering in the intervening layers traps 
photons in the shock and they are advected outward by the flow. When it reaches the surface of the star, the shock 
abruptly accelerates in the steep density gradient there, becoming even hotter and releasing a sharp burst of 
photons into the surrounding medium.  This results in an extremely bright breakout transient with peak bolometric 
luminosities $\gtrsim$ 10$^{45}$ erg s$^{-1}$, chiefly in the form of x-rays, hard UV and occasionally gamma rays.  
These radiative losses alter the hydrodynamics of the system and so cannot be ignored.  

The ambient medium into which the shock breaks out can strongly affect light curves and spectra in ways that 
may have been underestimated in past studies.   The luminosity and hardness of the breakout transient 
are governed in part by how much the shock accelerates as it exits the star:  the steeper the gradient, the 
brighter and harder the pulse.  Radiation breakout can occur after shock breakout in 
dense circumstellar envelopes because photons can be trapped by the wind after the shock exits the surface 
of the star \citep[][L.H. Frey et al., in preparation]{ofek10,bl11,chev11}.  Shock breakout has been the subject of numerous analytical studies 
\citep{cg74,mm99,ns10,Piro10,Katz11} and numerical simulations \citep{Ens92,Blinn00,Tomin09,fet09,Tolstov10,Fryer10,
kasen11} in the three decades since it was first predicted \citep{cg74,kl78}.  However, there have only recently been 
observations which are interpreted by some as shock breakout.  These include direct observations of X-ray pulses in 
GRB 060218/SN2006aj \citep{camp06} and XRO 080109/SN2008D \citep{sod08}, early-time peaks in UV light curves 
\citep{sch08,gez08}, and indirect observations through IR echoes in the Cas A SN remnant \citep{dwek08}.  

The location and time at which radiation breakout occurs depend on the radius at which the star or its 
environment becomes optically thin and is a function of wavelength.  Standard blackbody and light-crossing time 
arguments for peak breakout luminosities and pulse width assume a gray opacity and no attenuation beyond the $\tau$ = 1 
surface.  In reality, the opacity can be a strong function of wavelength in SN ejecta and is poorly approximated by a 
gray opacity.  Analyses of shock breakout observations \citep[i.e.][]{wax07,chev08,gez08,sod08} have 
considered a variety of progenitors and stellar environments but treat breakout as a wavelength-independent event 
that occurs at a single radius, temperature and time.  Depending on the wavelength band, this assumption can introduce 
significant errors to estimates of explosion properties derived from observations.  

The breakout transient is generally followed by a rise and then gradual decline in bolometric luminosity that is powered  
by both the conversion of kinetic energy into thermal energy and the decay of \Ni.  Shock heating from the supernova ejecta interacting 
with the surrounding winds also contributes to the peak luminosity in Type II SNe, and to a lesser extent in Ib/c SNe.  
Shocks in Type Ia SN surroundings can also contribute to the light curves near peak luminosity.  
This luminosity powered by shocks and \Ni\ decay persists 
for several months in core-collapse and Type Ia explosions and for 2 -- 3 years in pair-instability explosions. 
The fireball cools as it expands and the peak of its spectrum evolves to longer wavelengths. At the same time, the 
envelope surrounding the star that was ionized by the breakout pulse begins to recombine and blanket the 
spectrum with lines, particularly at shorter wavelengths.  In the frame of the shock, the photosphere from which
photons escape descends deeper into the ejecta as it expands.  The heavy elements it exposes to 
the surrounding medium over time depend on how Rayleigh-Taylor instabilities and asymmetries mix the interior of the 
star as the shock propagates to its surface.  When there is little mixing, lines due to heavier elements do not appear 
in the spectra until later times, but when mixing is extensive such lines can become manifest at early times.  The 
order in which metal lines appear in SN spectra over time can be a powerful probe of mixing during the 
explosion, and of the central engine itself.

Bolometric luminosities can surge at late times if the ejecta is dense and delays the escape of IR photons 
or if the photosphere moves into a warm region such as a \Ni\ layer.  Dense winds can also impact spectra at late times by  
more thoroughly blanketing the spectrum with lines, particularly at short wavelengths.  The light curve can 
also brighten if the ejecta collides with a dense shell from a violent luminous blue variable-type mass ejection (as in some Type IIn SNe).  
Such collisions can change even dim SNe into extremely bright events \citep[e.g.][]{nsmith07,vmarle10,
moriya10,Rom12}.  
After breakout, a radiative precursor, which would not be seen in hydro-only codes, propagates in front of the shock and 
interacts with any surrounding material.  If a shell was previously ejected, this precursor impacts the shell before the 
main shock and causes a bump in the light curve (D.J. Whalen et al., in preparation).  
During the nebular phase, the influence of shock heating and radiative losses on the hydrodynamics are minimal and hydro-only 
codes can be used for accurate simulations.  As departures from LTE become more significant during the nebular phase, 
the accuracy of the codes and opacities presented here will decrease.

We describe the LANL RHD code RAGE and how it is 
used to model supernova explosions in Section \ref{sec:rage}.  We review the LANL SPECTRUM code, which post-processes 
RAGE profiles to compute time-dependent spectra and light curves in Section \ref{sec:spectrum}, and 
discuss how atomic opacities derived from the LANL OPLIB database are implemented in both codes in Section \ref{sec:opac}. 
In Section \ref{sec:spec_tests} we present the suite of tests used to verify SPECTRUM.  We analyze a series of tests in 
which more physical processes are progressively activated in RAGE in order to assess their effects on spectra and light 
curves in Section \ref{sec:rage_tests}.  We show why full RHD and accurate 
opacities are so important to realistic supernova light curves and spectra in Section \ref{sec:discussion}, and in 
Section \ref{sec:conclusion} we summarize our conclusions and discuss future applications.

\section{RAGE} \label{sec:rage}

RAGE (Radiation Adaptive Grid Eulerian) is a multidimensional, multispecies Eulerian adaptive mesh 
refinement RHD code developed at LANL \citep{rage}.  It couples second-order conservative Godunov 
hydrodynamics to either gray or multi-group flux-limited diffusion radiation transport.  RAGE models radiation 
hydroflows on one-dimensional (1D) Cartesian, cylindrical and spherical polar coordinate meshes, two-dimensional (2D) Cartesian and cylindrical grids, 
and three-dimensional (3D) Cartesian boxes.  The mass fraction of each species is evolved with its own continuity equation under the 
assumption that all species have the same velocity at a given mesh point.

\subsection{Microphysics}

Although RAGE has three-temperature (electron, ion, and radiation) capability, we use two-temperature (2T) physics in 
our simulations, in which radiation and matter, while coupled, can be 
at different temperatures.  This can occur during shock breakout, when matter is known to be out of thermal 
equilibrium with photons \citep{ns10}.  This 2T physics is an important improvement over the pure 
hydrodynamics and one-temperature radiation models used in previous SN light-curve calculations. However, 
deviations between the matter and radiation temperatures can imply departures from local thermodynamic equilibrium 
(LTE) and lead to inaccuracies in our opacities during breakout, as we discuss below.  We do not explicitly 
transport gamma rays from the radioactive decay of \Ni\ in the ejecta.  Instead, their energy is deposited locally in 
the gas according to 
\vspace{0.05in}
\begin{equation}
\frac{dE}{dt} = \frac{E_{Ni}}{\tau_{Ni}}e^{-t/\tau_{Ni}} + \frac{E_{Co}}{\tau_{Co}-\tau_{Ni}}(e^{-t/\tau_{Co}}-e^{-t/\tau_{Ni}}),
\vspace{0.05in}
\end{equation}
where $\tau_{Ni}$ = 7.6$\times10^5$ s, $\tau_{Co}$ = 9.6$\times10^6$ s and the mean energies released per atom 
for the decay of \Ni\ and \Co\ are $E_{Ni}$ = 1.7 MeV and $E_{Co}$ = 3.67 MeV \citep{fet09}.  
This approximation is valid at early times when the ejecta is dense and optically thick but can break down at later 
times when gamma-rays are no longer trapped. This occurs after about 50 days in Type Ia SNe and, depending on the 
degree of mixing, after 150 days in core-collapse SNe. The energy deposition described above has been tested by comparing 
it to non-thermal $\gamma$-ray transport, which produces the same result at early times \citep{fet09}.  
We apply the same constant specific heat $C_v$ to each 
constituent material, so only ions contribute to the total pressure.  We also use an ideal gas equation of state 
(EOS) because the densities commonly found in our astrophysical applications are off of existing EOS tables in 
RAGE by many orders of magnitude.

\subsection{Radiation Transport}

Both gray and multi-group flux-limited diffusion are implemented in RAGE with opacities 
derived from the LANL OPLIB database using the TOPS code, as explained in Section \ref{sec:opac}.  Rosseland mean 
opacities are used to evaluate the diffusion coefficient in the transport equations and Planck mean opacities 
are used to calculate the emission and absorption terms.  All the tests in our study were done with gray flux-limited 
diffusion.  OPLIB opacities are calculated under the assumption that the gas is in LTE and are thus functions of 
frequency, temperature and density only. At late times, when the SN becomes transparent, LTE may no longer be valid 
because individual fluid elements become too diffuse to remain in thermal and radiation equilibrium with their 
surroundings.  When this happens, the true opacity can deviate from those in our tables.  The tests we describe in 
Section \ref{sec:rage_tests} show that LTE is a good approximation during shock breakout.  We also include an advection 
term in the gray diffusion calculation.  This advection term conserves energy (consistently incorporates
the work defined by pressure and change in volume, P*dV) and includes material motion effects (assuming the radiation flows with the matter).  Although this
is not correct in the free-streaming limit, this error is likely to be smaller than 
the fact that we are using flux-limited diffusion to approximate the transition 
from diffusion to free-streaming.  

\subsection{Gravity}
Gravity can play an important role in core-collapse supernovae 
at very early times, causing ejecta to fall back onto the newly formed compact remnant.  In most supernovae, the bulk 
of the fallback occurs in the first 1000 s of the explosion \citep{fcp99,fhh06,zwh08,fryer09}.  In our calculations, 
we first evolve SNe until after fallback and nuclear burning is complete in codes such as Kepler, CASTRO 
\citep{Almgren2010}, SNSPH \citep{frw06} and others \citep{fryer99}.  This 
is done in order to determine nucleosynthetic yields and how much material falls back onto the compact remnant 
so this material can be removed from the profiles prior to mapping them into RAGE.  After fallback and burning is 
complete, gravity no longer plays a large role in the explosion and we neglect it in RAGE.  As a consequence, the 
star can slightly expand into the surrounding medium in the time it takes the shock to reach the surface, but this 
motion is minor.  If we regrid the explosion prior to breakout, the original profile of the star is restored to the 
grid.

\subsection{Supernova Profiles}

Gas densities, velocities, specific internal energies (erg g$^{-1}$), radiation energy densities (erg cm$^{-3}$) 
and species mass fractions are required to initialize an explosion profile in RAGE. Gas densities, mass fractions and 
velocities are directedly imported from codes 
such as those listed above.  However, rather than converting the gas energies in those codes into the specific 
internal energies required by RAGE, we calculate gas and radiation energies from the temperature of the gas, 
assumed to be ideal and monatomic, which is also usually supplied by the codes:
\vspace{0.05in}
\begin{equation}
E_{gas} = \frac{3}{2}RT \vspace{0.05in}
\end{equation}
and
\vspace{0.05in}
\begin{equation}
E_{rad} = aT^4, \vspace{0.05in}
\end{equation}
where $R$ is the ideal gas constant and $a$ is the radiation constant.  This avoids subtleties in how gas energies are 
defined in various codes, some of which lump contributions due to the ionization states of atoms and radiation in with 
the internal energy.  In contrast to internal energy, the temperature is unambiguously defined in most codes.  

We map the profile, which consists of the explosion, the surrounding star and the circumstellar environment, onto 
a 1D spherical grid in RAGE.  We choose the grid resolution so that the blast and its photosphere 
are well resolved, and we permit up to five levels of refinement in both the initial interpolation of the profile onto 
the grid and during the simulation.  We set reflecting and outflow boundary conditions on fluid and radiation flows 
at the inner and outer boundaries of the mesh, respectively.  When the calculation is begun, Courant times are 
initially small due to high temperatures, large velocities and small cell sizes.  To reduce execution times and 
accommodate the expansion of the ejecta, we periodically regrid the profiles onto a larger mesh as the explosion 
grows.  This significantly increases the time step on which the model evolves.  We remap just the explosion itself, 
ignoring any medium beyond the shock or its radiation front, and then graft the original environment lying beyond 
this radius back onto the shock or front on the new grid.  We apply the same criteria in choosing a new grid 
as in the original setup.  The entire calculation typically requires 20,000 CPU hours on LANL platforms.

\section{SPECTRUM} \label{sec:spectrum}

\begin{figure*}
\epsscale{1.0}
\plotone{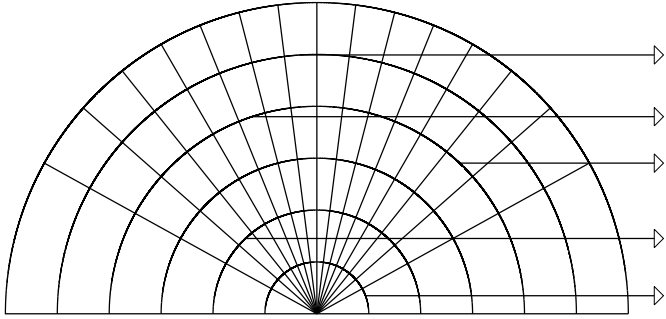}
\caption{The SPECTRUM mesh. The radius of a given zone is defined at its lower face but $\mu = cos\, \theta$ is 
cell-centered. The grid is uniformly partitioned in $\mu$.} 
\vspace{0.1in}
\label{fig:grid1}
\end{figure*}

We post-process RAGE hydrodynamical profiles with the LANL SPECTRUM code to construct light curves and 
spectra for our simulations.  Stand-alone post-processing allows us to exploit the comprehensive LANL OPLIB 
opacity database to compute detailed absorption and emission features in our spectra without resorting to thousands 
of energy groups in the transport calculations. These monochromatic opacities are calculated assuming LTE conditions for a 
range of temperatures, densities and materials; this database is described in detail in Section \ref{sec:opac}.  
SPECTRUM is general enough that it can accommodate hydrodynamic 
profiles from any code.  The current version of SPECTRUM is parallelized and reads in pre-computed 
opacity tables, derived from the LANL OPLIB database, unlike the earlier version summarized in 
\citealt{fet09}.  This previous version was serial, did not efficiently reference the frequency grid and required 
an additional external calculation to determine the opacities for each radial grid point.  

The calculations described below, with the monochromatic OPLIB opacities in the spectrum calculations, include the 
effects of line and 
continuum opacities, fluorescence, Doppler shifting, time dilation and limb darkening.  Other ways of calculating these 
effects, such as the Sobolev approximation, are not necessary since we include them explicitly.  

For 1D RAGE simulations, densities, velocities, mass fractions and radiation temperatures from the finest level of 
refinement of the adaptive mesh grid are extracted and ordered by radius.  These profiles typically contain more than 
50,000 radial data points, but constraints on machine memory and time prevent us from using all of them in a 
SPECTRUM calculation.  A subset of the data from each RAGE profile must be carefully chosen to fully resolve all 
radiating regions of the flow from which photons can escape, such as the photosphere of an SN shock or the collision 
of a shock with a dense structure.  The radial grid used in SPECTRUM for one set of SN simulations is described in 
more detail in Section \ref{sec:rad_bin}.  

\subsection{Luminosity of a Fluid Element}
The 1D RAGE data, sampled as described above, is mapped onto a 2D grid of radial and angular bins in SPECTRUM 
as shown in Figure \ref{fig:grid1}, where the angular bins are uniform in $\mu = cos\, \theta$.  
SPECTRUM takes the fluid at each mesh point to be in local thermodynamic equilibrium (LTE) with its surroundings and 
its radiation spectrum to be Planckian.  In this approximation, each fluid element emits radiation at the same rate it 
absorbs it from its environment, so the Kirchhoff-Planck relation holds:
\vspace{0.05in}
\begin{equation}
\eta \, = \, \kappa_{abs} I \, = \kappa_{abs} B_{\nu}, \vspace{0.05in} \label{eq:KP}
\end{equation}
where $\eta$ is the emissivity of the fluid parcel, $\kappa_{abs}$ is its monochromatic absorption opacity at
frequency $\nu$ and \vspace{0.05in}
\begin{equation}
B_{\nu} \, = \, \frac{2h{\nu}^3}{c^2} \frac{1}{e^{h\nu/kT_{gas}} - 1} . \vspace{0.05in} \label{eq:bnu}
\end{equation}
From Equations \ref{eq:KP} and \ref{eq:bnu} we construct the luminosity of each fluid element into a given line of 
sight: \vspace{0.1in}
\begin{equation}
L_{\nu}^{r, \mu}  = \kappa_{abs} m \frac{2h{\nu}^3}{c^2} \frac{1}{e^{h\nu/kT_{gas}} - 1} \frac{(1-v\mu/c)^2}{\sqrt{(1-(v/c)^2)}}. \vspace{0.1in}
\label{eq:lnu}
\end{equation}
Here, $r$ and $\mu$ define the radial and angular position of the fluid parcel, $m, v$ and $T_{gas}$ are its mass, 
velocity and temperature, $\nu$ is the frequency at which it radiates in its own frame, and $h, c$ and $k$ are 
Planck's constant, the speed of light and Boltzmann's constant, respectively.  Note that the spherical symmetry of 
the problem implies that the luminosity defined in Equation \ref{eq:lnu} is for the entire ring of material passing 
through the given mesh point in Figure \ref{fig:grid1} and centered on the $\mu=1$ axis.   Consequently, $m$ is the 
mass of the ring: 
\vspace{0.05in}
\begin{equation}
m_i \, = \, \frac{4}{3}\frac{\pi}{n_A}(r_{i+1}^3 - r_{i}^3) \rho_{i},   \vspace{0.05in}
\end{equation}
where $\rho_i$ is the ejecta density at $r_i$ and $n_A$ is number of angular bins.

Discretizing the spherically symmetric blast profile uniformly in $\mu$ guarantees that every fluid element at a 
given radius has the same volume and so the same luminosity.  The $v/c$ term in the denominator adjusts the luminosity 
to account for the relativistic time dilation associated with the motion of the fluid and the $v/c$ term in the 
numerator accounts for Doppler shifts in photon energy due to motion relative to the line of sight.  As shown in 
Figure \ref{fig:grid1}, if the ejecta is expanding photons from $\mu <$ 0 are redshifted and from $\mu >$ 0 are 
blueshifted. 

\subsection{Escape Luminosity}

\begin{figure}
\epsscale{1.17}
\plotone{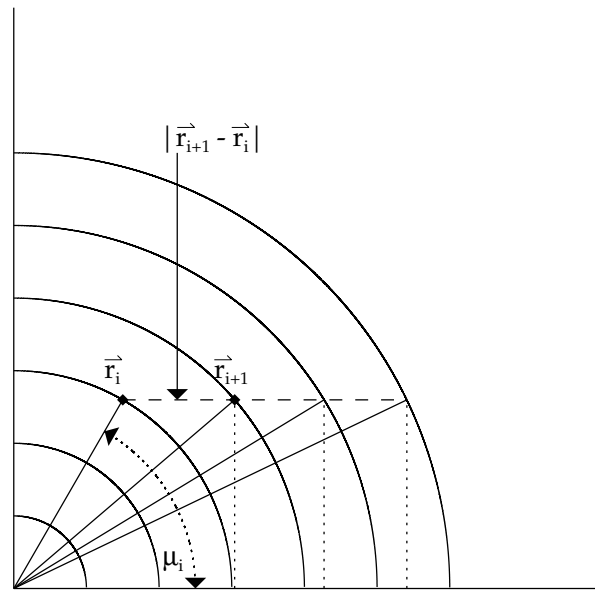}
\caption{The paths over which the $\tau_{r_i, \mu_j}$ are calculated.  The $r_i$ and $\mu_j$ of a given fluid 
element set the lengths of all segments from its position out to the edge of the grid along the line of sight since 
the $\left|r\right|$ are simply the radii of adjacent mesh points and the ends of the segments all lie at the same 
height above the axis of symmetry.}
\label{fig:grid2}
\end{figure}

To calculate the luminosity from each fluid parcel that actually escapes into the interstellar medium we attenuate 
$L_{\nu}^{r_i, \mu_j}$ by $e^{-\tau_{r_i, \mu_j}}$, where $\tau_{r_i, \mu_j}$ is the integrated optical depth from the 
mesh point out of the ejecta to photons at frequency $\nu$.  It is constructed from consecutive line segments as shown 
in Figure 
\ref{fig:grid2}: 
\vspace{0.1in}
\begin{equation}
\tau_{r_i, \mu_j} = \sum_{i} \kappa_{abs,i} \rho_i \left|{\bf{r}}_{i+1} - {\bf{r}}_i\right|, \vspace{0.02in}
\label{eq:atten}
\end{equation}
where $\rho_i$ is the density along the segment (assumed to be that of the mesh point), and the sum is over all 
line segments extending from the source to the outer edge of the grid along the line of sight.  As shown in Figure 
\ref{fig:grid2}, the $r_i$ and $\mu_j$ of the fluid parcel uniquely determine the lengths of all segments from its 
position out to the edge of the grid because all the $r$ are known and the ends of each segment are at the same 
height above the axis of symmetry.  Each zone along the line of sight from the radiating cell intercepts photons at 
an energy different from that at the source that is determined by the total relative Doppler shift between them.  
Thus, the $\kappa_{abs}$ for each line segment is computed at the energy into which source photons are Doppler 
shifted upon reaching that zone. 

\begin{figure}
\epsscale{1.17}
\plotone{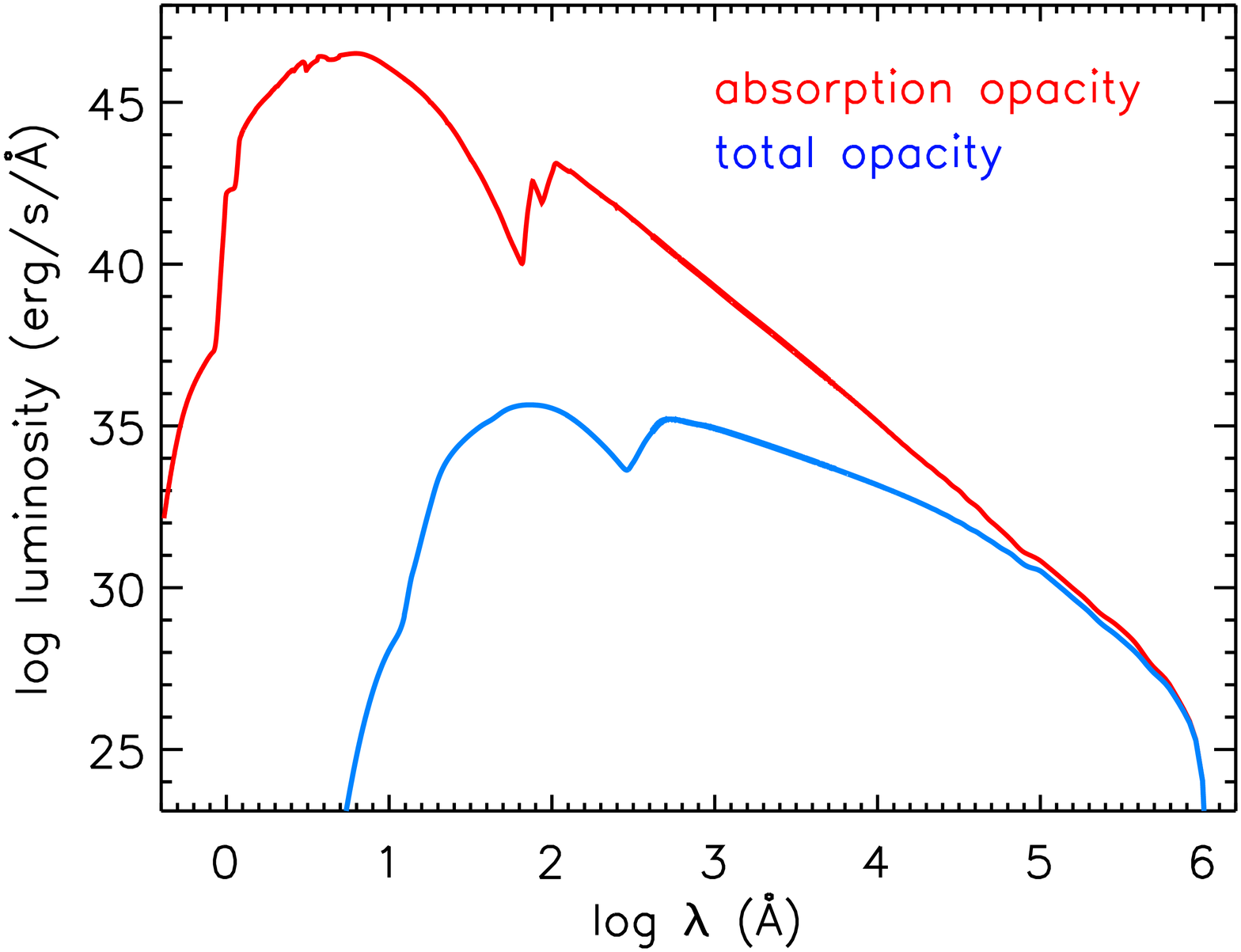}
\caption{Spectra from a 225 M$_{\odot}$ Pop III pair-instability SN at 1.3 hours, calculated with either 
$\kappa_{tot}$ or $\kappa_{abs}$ in Equation \ref{eq:atten}.  Note that luminosity is preferentially lost at short 
wavelengths.}
\label{fig:comp_opac} \vspace{0.2in}
\end{figure}

Since Equation \ref{eq:atten} is used to calculate the contribution to the escaping luminosity, the use of 
$\kappa_{abs}$ here is appropriate as only the photons that are actually destroyed do not escape.  However, in 
situations where there is a strongly scattering dominated atmosphere with little or no absorption, the thermalization 
depth of the photons will be significantly different from the depth of last interaction.  While these scattered 
photons are not destroyed, their properties may change due to a scattering event and their emergence from 
the ejecta will be undoubtedly delayed in time.  The approximation in Equation \ref{eq:atten} will remain 
adequate for modeling the emergent spectrum so long as two conditions are met: (1) the added delay from scattering 
processes near the thermalization surface ($\tau_{abs} \sim 1$)  is short compared to the ejecta evolution timescale, 
and (2) the spectral information in the emerging photons is not strongly altered by the scattering interactions.  
For typical conditions in low energy X-ray emission from supernova ejecta, Thompson electron scattering is sufficient 
and its coherent nature assures that the second criteria is met.  In reality Thompson scattering limit is not 
perfectly coherent and attention must be paid to the number of scatters to ensure that the integrated energy shift is 
not significant.  
The first criteria is one that must be checked for each new ejecta scenario and is an important caveat in the use of 
the SPECTRUM code.  Scattering dominated atmospheres can occur in extremely energetic explosions, such as the 
pair-instability SNe of blue compact giant stars \citep{wet12a}.  In these cases, the flux calculated in SPECTRUM 
with $\kappa_{tot}$ in Equation \ref{eq:atten} instead of $\kappa_{abs}$ can be 10 orders of magnitude lower, with 
a peak wavelength 10 times longer, as shown in Figure \ref{fig:comp_opac}.  

Due to the high velocities seen at the shock front, even with well-resolved data there can be a significant velocity 
gradient across just one zone. Consequently, photons from one end of the zone can be removed at a different energy 
at the other end that is determined by the difference 
in velocity between the two ends.  Depending on the source frequency $\nu$, several energy bins may span the 
Doppler shift of the velocity gradient across the zone.  We therefore frequency average the opacity of every zone by 
first computing its total opacity at each energy bin bracketed by the Doppler shift.  We then sum them and divide by 
the number of bins across the shift to obtain the average opacity of the zone, effectively giving us subgrid resolution 
for our opacities.  The total luminosity escaping the 
explosion along the line of sight at source frequency $\nu$ is then just the sum of the luminosities of each mesh 
point attenuated by the matter along the line of sight
\begin{equation}
L_{\nu}^{tot} = \sum^{i=n_R}_{r=1} \sum^{j=n_A}_{j=1} L_{\nu}^{r_i, \mu_j} e^{-\tau_{r_i, \mu_j}},
\label{eq:lnutau}
\end{equation}
where $n_R$ and $n_A$ are the number of radial and angular bins, respectively.  
Equations \ref{eq:lnu} and \ref{eq:lnutau} together with the symmetry of the grid ensure that limb darkening, 
Doppler shifts and time dilation due to relativistic expansion of the ejecta are included in our spectrum 
calculations.  We also account for atomic emission lines with the $\kappa_{abs}$ term in Equation \ref{eq:lnu} 
and absorption lines with the $e^{-\tau_{r_i, \mu_j}}$ term in Equation \ref{eq:lnutau}.  Note that departures from 
LTE in the ejecta can lead to errors in the SPECTRUM opacities, and hence luminosities.  Furthermore, deviations from 
LTE may also partially invalidate Equation \ref{eq:KP} and cause additional errors in our spectra, especially at late 
times.

The 14,900 energy bins for which there are opacities in the LANL OPLIB database, described in Section \ref{sec:tops_opac}, 
can change from point to point on the grid because the profile temperature varies with radius.  
These bins range from 0.8 \AA\ -- 2.0 $\times$ 10$^7$ \AA\ at 0.5 eV to 4.0 $\times$ 10$^{-6}$ \AA\ -- 100 \AA\ at 
10$^5$ eV. To calculate the $L_{\nu}^{r_i, \mu_j}$ one must choose a fixed set of 14,900 $\nu$ for all the spectra.  
The bins we adopt are those for OPLIB opacities at $T = $ 1 eV.  These energies correspond to wavelengths of 
0.415 to 9.1 $\times$ 10$^{6}$ \AA, which capture all spectral features from shock breakout to the end of the light 
curve.  We calculate $L_{\nu}^{r_i, \mu_j}$ using the energy grid for the temperature in that cell and then map 
each $L_{\nu}$ into the standard $T = $ 1 eV energy grid.  

The SPECTRUM code can be extended to post-process data from 2D RHD simulations.  The 
luminosity and the attenuation along a line of sight are calculated with the same methods described by Equations 
\ref{eq:lnu} and \ref{eq:atten}. The 2D discretization is done by taking parallel slices through the grid that align 
with the desired line of sight. Each slice can then be solved as separate 1D spectra calculation and the total 
luminosity can be obtained by summing the spectra of individual slices. How we extend SPECTRUM to general 2D data sets 
is described in detail in (W. Even et al., in preparation).

\subsection{Light Curves}

Bolometric light curves can be constructed with SPECTRUM by summing over the full energy range of the spectra at each 
time step.   Light curves for a wavelength band from a specific instrument can be created by convolving the spectra and a 
filter response function.  Each 
SPECTRUM calculation averages 200 CPU hours so the total CPU time required to compute a 250-point light curve is one 
to two times that of the RAGE run.  However, because many SPECTRUM jobs can run in simultaneously and each one is 
completed quickly, it is usually possible to calculate a light curve per day on LANL platforms.  

\section{Opacities} \label{sec:opac}

The Rosseland and Planck mean opacities in RAGE and the monochromatic opacities in SPECTRUM are calculated 
with the LANL TOPS code\footnote{http://aphysics2/www.t4.lanl.gov/cgi-bin/opacity/tops.pl/} from cross sections compiled in the 
OPLIB database \citep{oplib}.  An advantage to using this database is that all of the opacities are calculated in a 
consistent manner, i.e. the results have not been culled from a variety of sources.  A disadvantage is that only 
electric dipole transitions are considered in the bound-bound contributions, while other databases include forbidden 
lines.  However, for LTE conditions, E1-allowed transitions should be sufficient for capturing the relevant physics.  
We also note that the OPLIB database averages together some nearby spectral lines, but this level of 
accuracy is sufficient for the light-curve modeling considered here as well as for identifying general spectral 
features.  

We now describe these cross sections and opacities in detail and how mixed opacities 
are calculated for multispecies flows in RAGE and SPECTRUM.

%
\begin{deluxetable}{llllc} 
\tabletypesize{\scriptsize}  
\tablecaption{OPLIB Opacity Frequency Grid  \label{table:t1}}
\tablehead{
\colhead{ $\Delta u$} & \colhead{} & \colhead{$u$ range} & \colhead{} & \colhead{\# of grid points}}
\startdata 
0.00125  & \hspace{0.4in}   &  0.00125 -- 12       &  \hspace{0.4in}    &   \hspace{0.1in} 9600    \\
0.005      &     &  12.005 -- 20         &      &   \hspace{0.1in}  1600   \\
0.01         &    &  20.01 -- 30           &       &   \hspace{0.1in}  1000   \\
0.1           &     &  30.1 -- 100           &      &   \hspace{0.1in}  700      \\
1              &     &  101 -- 1000          &      &   \hspace{0.1in}  900      \\
10            &     &  1010 -- 10000     &      &   \hspace{0.1in} 900       \\
100          &     &  10100 -- 30000  &       &   \hspace{0.1in}  200     \\
& \vspace{-0.05in}& & & 
\enddata
\label{tab:ubins}
\end{deluxetable}  
\vspace{0.2in}

\subsection{The LANL OPLIB Database}\label{sec:tops_opac}

TOPS can calculate external tables of absorption and scattering opacities $\kappa$ for a set of density and 
temperature points that the user chooses.  The $\kappa$ are derived from cross sections $\sigma$ in the 
OPLIB database that have been compiled for pure chemical elements that are assumed to be in LTE, according to 
\vspace{0.05in}
\begin{equation}
\kappa_{\nu}(\rho,T) = \frac{N_{0}}{M}\sigma_{\nu}(\eta,T),\vspace{0.05in} \label{eq:single_opac}
\end{equation}
where $N_{0}$ is Avogadro's number, $M$ is the atomic weight of the element and $\rho$ is the density. 
The $\sigma$ are calculated from first principles based on quantum mechanics, and are tabulated on a grid of 
temperatures $T$, electron degeneracies $\eta$, and dimensionless reduced photon energies $u$ given by  \vspace{0.05in}
\begin{equation}
u \equiv \frac{h\nu}{kT}. \vspace{0.05in}
\label{eq:ugrid}
\end{equation}
This (T,$\eta$,u) grid is fixed for all elements.  
For a given $\eta$ and $T$, there are 14,900 $u$ that range from 1.25 $\times$ 10$^{-3}$ to 3 $\times$ 10$^4$, 
which we summarize in Table \ref{tab:ubins}.  

\begin{figure}
\epsscale{1.17}
\plotone{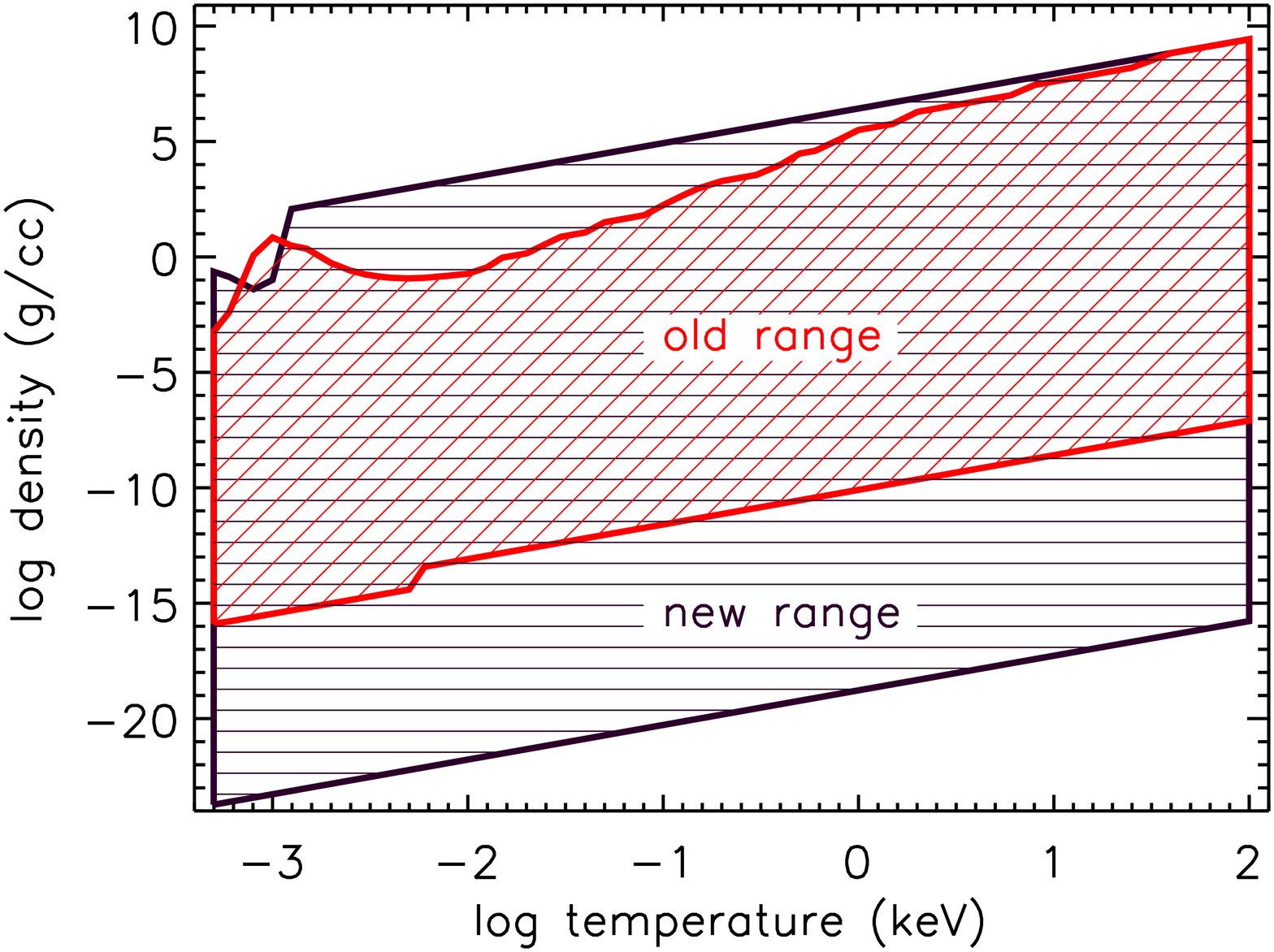}
\caption{The densities and temperatures for which hydrogen cross sections exist in the LANL OPLIB database. In 
general, the density range varies with both temperature and material.  The red hatched lines indicate the 
original density range of the hydrogen cross sections and the black lines delineate our extension to this data set.}
\label{fig:topslims}
\vspace{0.1in}
\end{figure}

TOPS can calculate gray, multi-group, or monochromatic opacities for pure elements or mixtures of 
elements and chemical compounds.  Opacities in RAGE and SPECTRUM are referenced by $\rho$ and $T$ but, 
as mentioned above, cross sections are tabulated by $\eta$ and $T$ in OPLIB.  TOPS determines the $\rho$ 
and $T$ associated with an $\eta$ and $T$ according to the standard relation \vspace{0.1in}
\begin{equation}
\rho(\eta,T) = \frac{MN_{I}(\eta,T)}{N_{0}} = \frac{MN_{e}(\eta,T)}{\bar{Z}(\eta,T)N_{0}},  \vspace{0.05in}
\label{eq:topsrho}
\end{equation}
where $\rho$ is the density of the pure element or mixture, $N_I$ is its ion number density and $\bar{Z}$ is its 
average charge.  The relation between $N_e$ and $\eta$ is assumed to be that of a Fermi degenerate ideal gas:
\vspace{0.05in}
\begin{equation}
N_e = \frac{4}{\sqrt{\pi}}\frac{(2\pi m_e kT)^{3/2}}{h^3} F_{1/2}(\eta), \vspace{0.05in}
\label{eq:eta}
\end{equation}
where $F_{1/2}(\eta)$ is the Fermi-Dirac integral of order $1/2$, $k$ is the Boltzmann constant, $h$ is Planck's 
constant and $m_e$ is the mass of an electron.
When constructing an opacity table, TOPS will produce data only at temperatures that exist in the OPLIB database, 
i.e. neither interpolation nor extrapolation are performed for requested T that are not in the database.  For this 
reason, the $T$ points we adopt in our opacity tables are the same as those in OPLIB.  On the other hand, if a 
requested density point falls between two corresponding $\rho$ for which OPLIB $\sigma$ exist, TOPS will interpolate 
the cross section and compute an opacity for that point.  If the density point falls above or below the corresponding 
range of $\rho$ for which $\sigma$ are tabulated, TOPS returns the opacity of the closest corresponding 
on-table $\rho$ and does not extrapolate.  This procedure can lead to serious errors in opacity at very high or low 
densities that are beyond the OPLIB limits.  The astrophysical simulations for which SPECTRUM was developed frequently 
have densities that are off-table on the low end of the original OPLIB grid.  We therefore extended the database 
downward by 10 orders of magnitude in density for H, He, C, N and O, as shown in Figure \ref{fig:topslims} for H.  
The region in $\rho$ and $T$ for which opacities of all the elements can be calculated is very similar to the red 
region shown for H; the extended regions for He, C, N and O are very similar to the black region showing our extension to 
OPLIB for H. The TOPS opacity tables are output to files that are read in by RAGE and SPECTRUM at execution time.  

\begin{figure*}
\plotone{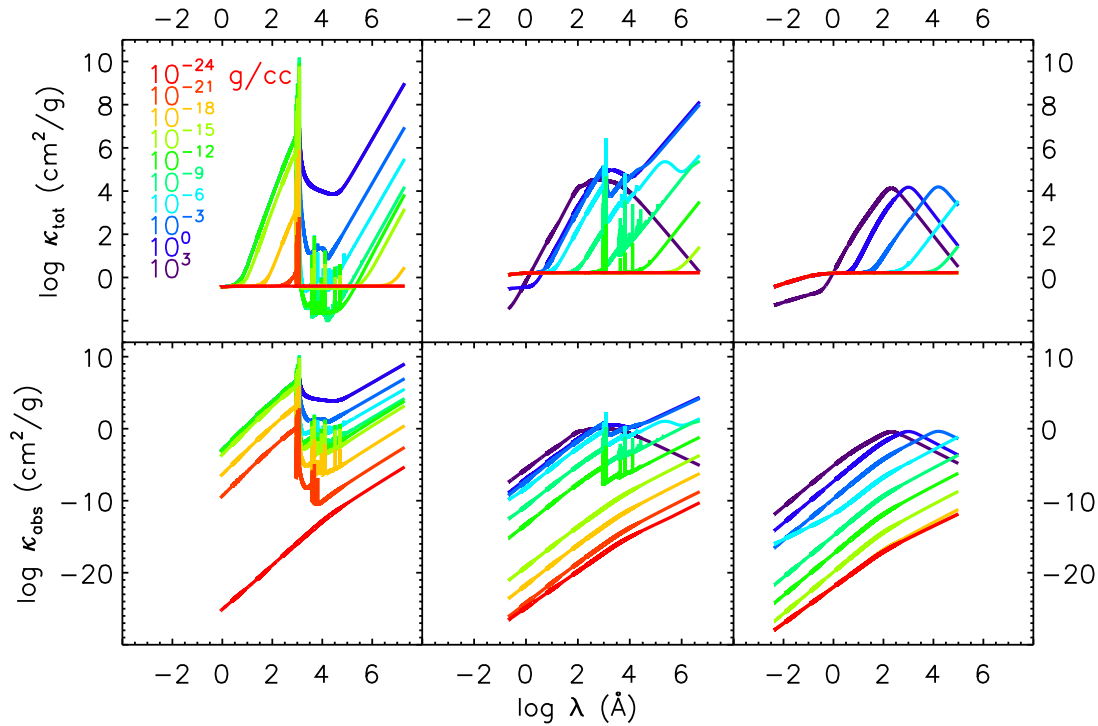}
\caption{Total opacity (upper row) and absorption opacity (lower row) for hydrogen as a function of 
temperature and wavelength.  Left to right: $T =$ 0.5 eV, 2.0 eV, and 100 eV.}
\label{fig:H_opac}
\end{figure*}

We show how opacities for hydrogen vary with density and temperature in Figure \ref{fig:H_opac}.  Absorption opacities 
at any given density generally fall as the temperature rises and the atoms become increasingly
ionized (note that the total opacity falls to the Thomson scattering floor at sufficiently short wavelengths).  
Also, the absorption lines that are so prominent at lower temperatures fade at higher ones as the number of 
bound electrons decreases in the increasingly ionized gas.  These trends demonstrate that the atomic physics included 
in the OPLIB database captures the effect of temperature and density on opacity.

In certain applications it is advantageous to create tables of opacities for a multi-material mixture with TOPS.  To 
calculate mixed opacities, TOPS assumes that all constituents are homogeneously mixed in each cell (i.e. atomically 
mixed) and imposes the EOS condition of temperature-$\eta$ (T-$\eta$) equilibrium on the constituents of the mixture.  
By virtue of Equation \ref{eq:eta}, imposing T-$\eta$ equilibrium is equivalent to imposing T-$N_e$ equilibrium, which 
is consistent with the physical picture of each constituent experiencing the same global free-electron distribution.  
Furthermore, since an EOS provides the relation between three physical quantities, imposing T-$N_e$ equilibrium among 
the constituents is equivalent to imposing T-$P_e$ equilibrium, i.e. the temperature and electron pressures associated 
with each constituent are equal to those of the mixture.  Under these assumptions, TOPS obtains the opacity for a 
mixture at conditions $\rho$ and T by summing the cross sections for pure elements according to \vspace{0.05in}
\begin{equation}
\kappa_{\nu}(\rho,T) = \frac{N_{0}}{M}\sum\limits_{i}f_{i}\sigma_{\nu}^{i}(\eta,T),\vspace{0.05in}
\label{eq:topsopac}
\end{equation}
where $N_{0}$ is Avogadro's number, $M$ is the atomic weight of the mix, $f_{i}$ is the atomic fraction of 
the $i^{th}$ element, and $\eta$ is chosen to satisfy Equation \ref{eq:topsrho}.  TOPS calculates gray or multi-group 
mixed opacities by integrating these monochromatic opacities with the appropriate weighting function.  

\subsection{RAGE Opacities} \label{sec:rage_opac}

We use TOPS to precompile Rosseland and Planck mean opacities for each pure element of the SN ejecta 
in a single external file for RAGE.  However, since a mesh point on the RAGE grid might be populated by 
several elements, RAGE must construct a single opacity for the mixture.  RAGE does not construct mixed 
opacities in the same way as TOPS because it imposes different EOS mix conditions on the constituents.  
Instead of taking them to be atomically mixed in each cell, RAGE assumes that they remain physically 
segregated (i.e. consistent with a heterogeneous or chunk mix), each with its own volume fraction and 
partial density \citep[for additional details concerning the atomic and chunk mix assumptions, 
see][]{Smith03,Pom91}.  Like TOPS, RAGE imposes T-$P$ equilibrium on the components of the mixture, but 
considers both the ion and electron pressure.  More specifically, in RAGE the ion pressure of each 
constituent is equal to the ion pressure of the mixture and the electron pressure of each component is 
the same as the electron pressure of the mixture.  After obtaining the partial densities from the above 
procedure, the opacity of the mixture is obtained from 
%
\begin{equation}
\bar{\kappa}(\rho,T) = \sum\limits_{i}f_{i}^{M}\bar{\kappa}_{i}(\rho_{i},T), \vspace{0.03in}
\label{eq:rageopac}
\end{equation}
where $\bar{\kappa}$ represents a mean (gray or group) opacity, $f_{i}^{M}$ is the mass fraction and $\rho_i$ 
the partial density of the $i^{th}$ constituent, which is calculated from its volume fraction and mass fraction.  
The RAGE simulations presented in this paper use gray opacities, so $\bar{\kappa}$ represents the 
Rosseland or Planck mean opacity.  In regions where ion pressure is non-negligible and material is not atomically 
mixed, Equation \ref{eq:rageopac} is expected to yield mixed opacities that are more accurate than those obtained 
under the assumption that electrons dominate the pressure, as in TOPS \citep{tops06}.  However, because we are using 
an ideal gas EOS rather than the RAGE tabular EOS, we ignore electron pressure when calculating 
the $\rho_i$. 

In principle, one should first construct the sum in Equation \ref{eq:rageopac} at each frequency and then
average over frequency to compute the mean mixed opacity rather than sum opacities that have already
been frequency-averaged.  In the Planck mean, the two approaches yield the same opacity because the 
mean is a simple arithmetic average for which summing and averaging are commutative. The two 
operations are not commutative for Rosseland opacities because the mean is harmonic, not arithmetic.  
Unlike TOPS, which frequency averages mixed monochromatic opacities to compute mean mixed opacities, 
RAGE sums mean opacities, even though it is somewhat less accurate, because computational constraints 
preclude a RAGE calculation for all 14,900 $u$.  Finally, to construct the mixed opacity at a given 
mesh point at some $\rho$ and $T$, RAGE performs a bi-linear interpolation of log $\bar{\kappa}$ from 
Equation \ref{eq:rageopac} over log $\rho$ and log $T$.  

\subsection{SPECTRUM Opacities} \label{sec:spectrum_opac}

We use TOPS to create monochromatic absorption and total opacity tables for SPECTRUM, one for each element.  
We calculated tables for the first 30 elements at the 50 OPLIB temperatures from 
$5 \times 10^{-4}$ keV to 100 keV and a grid of 81 densities from $10^{-15}$ to $10^5$ g/cm$^3$.  Opacities for H, 
He, C, N, and O were calculated for an additional 20 densities from $10^{-25}$ to $10^{-15}$ g/cm$^3$, using data 
from the extended tables described earlier.  Opacities for all 14,900 $u$ are tabulated for each $\rho$ and $T$.  The 
OPLIB $u$ grid structure (shown in Table 1) simplifies retrieval of the desired opacity from the tables because 
after the $u$ interval in which a given frequency $\nu$ resides has been determined, a simple index that directly 
references that interval is easily computed.  This approach results in spectrum calculation times that are much 
shorter than those done with bisection searches.

To calculate the mixed opacity for a cell at a given $\rho$ and $T$, SPECTRUM first computes the opacity 
$\kappa_i$ of each constituent of the mixture at its partial density $\rho_i$.   At the given frequency, log 
$\kappa_{\nu}^{i}$ is extracted from the table at the two densities that bracket $\rho_i$ at the table temperature 
closest to $T$.  
SPECTRUM then performs a simple linear interpolation of the logarithm of these two opacities with respect 
to log $\rho_i$ while holding $T$ constant to obtain the desired $\kappa_{\nu}^{i}$, i.e. as in TOPS, SPECTRUM does 
not interpolate $\kappa_{\nu}^{i}$ in $T$.  Like RAGE, the mixed opacity of the cell is then taken to be the sum of 
the opacity of each constituent weighted by its mass fraction, according to 

%
\begin{equation}
\kappa_{\nu}(\rho,T) = \sum\limits_{i}f_{i}^{M}\kappa_{\nu}^{i}(\rho_{i},T), 
\label{eq:specopac}
\end{equation}
which is the monochromatic version of Equation \ref{eq:rageopac}.

\section{SPECTRUM Tests} \label{sec:spec_tests}
We ran a series of tests to verify the SPECTRUM code and to determine the number and structure of radial bins and the 
number of angular bins required to converge to accurate spectra given memory and time constraints.  We find that 
convergence is problem specific, and should be investigated for each new type of simulation.  

\subsection{Blackbody Tests}

\begin{figure}
\epsscale{1.17}
\plotone{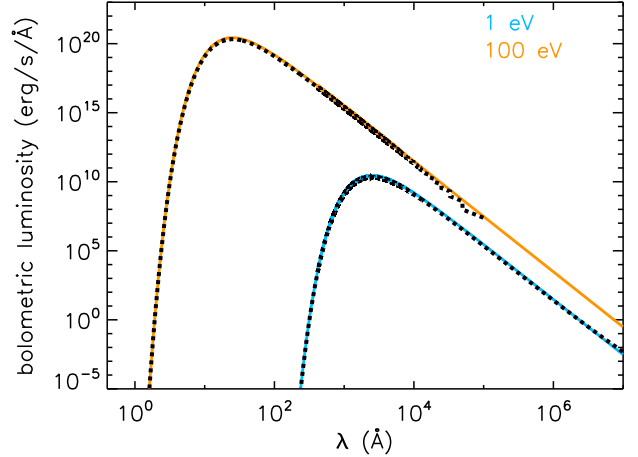}
\caption{Analytical (solid lines) and SPECTRUM (dotted lines) blackbody spectra at 1 eV and 100 eV.} 
\vspace{0.1in}
\label{fig:bb}
\end{figure}

One basic test every spectra code should be able to pass is to reproduce an analytic blackbody in the case of a 
constant opacity.  To do this, we calculated spectra for a static sphere with constant temperature, density and 
opacity, surrounded by a vacuum.  To obtain accurate blackbody spectra, the opacity and density must be selected so 
that the $\tau=$20 surface is located a small distance inside the outer surface.  We initialized the sphere in 
SPECTRUM with a very fine radial resolution at the emitting surface.  As we show in Figure \ref{fig:bb}, SPECTRUM 
accurately reproduces blackbody spectra for a range of temperatures.

\subsection{Convergence Tests}\label{sec:conv}

As explained in Section \ref{sec:spectrum}, for 1D problems SPECTRUM maps data onto a 2D grid with radial and angular 
bins.  We test the convergence of our SPECTRUM calculation in both angle and radius with spectra from a fiducial 
23 M$_{\odot}$ supernova.  The progenitor, which was evolved in TYCHO \citep{ya05}, is a binary star that loses its H 
envelope at the base of the red giant branch, evolves as a Wolf-Rayet star, and then ends its life as a 6.4 
M$_{\odot}$ CNO core that explodes as a Type Ic SN \citep{y06}. We first modeled the collapse of its core, the 
launch of the shock and all nuclear burning in a 1D hydrocode \citep{fryer99} and then mapped its profile into RAGE 
together with a wind envelope, defined with a mass loss rate of $10^{-5}$ M$_{\odot}$/yr and a velocity of 10$^8$ 
cm/s.  This star was originally investigated as a potential Cas A progenitor and is now being studied to understand 
the effects of pre-core collapse mass loss on shock breakout and SN light curves.  The convergence tests presented 
here are for a single time after shock breakout, 400 s, but we have performed identical tests at a variety of times 
before and after breakout and find similar results.  These radial and angular bin convergence tests were conducted in 
tandem over a 2D parameter space, 500-9000 radial bins and 2-400 angular bins, to find the best combination of radial and 
angular resolution.  

\subsubsection{Radial Convergence}\label{sec:rad_bin}

\begin{figure}
\epsscale{1.17}
\plotone{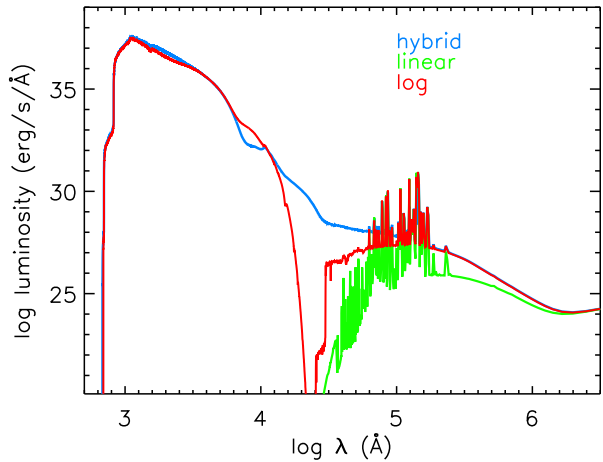}
\caption{
Output spectra for identical RAGE data from a Type Ic SN at 400 seconds after core-collapse, calculated with 160 angular bins 
and 5200 radial bins organized by three different binning schemes: linear bins, log bins, and the hybrid scheme described in 
the text.  
}
\vspace{0.1in}
\label{fig:comp_bin}
\end{figure}

As noted in Section \ref{sec:spectrum}, the RAGE simulations used as input to SPECTRUM contain at least 50,000 radial 
data points but current memory constraints limit SPECTRUM input to less than 10,000 points, requiring the creation of 
a new radial grid.  Densities, mass fractions, velocities, and temperatures from the finest mesh are extracted from 
the RAGE profiles and sequentially ordered by radius.  Three different binning schemes were considered: linear, logarithmic, 
and a hybrid scheme which focuses resolution on the radiating region.  
A simple binning scheme with uniform, linearly spaced, bins will completely exclude the interior of the simulation and 
only resolve the wind.  Logarithmically spaced bins include the full range of radii but do not sufficiently resolve 
the radiating region around the shock.  Figure \ref{fig:comp_bin} shows spectra for these two binning schemes.  To 
obtain an accurate binning scheme, any region where rapid changes in density, velocity or 
temperature occur and from which photons can escape, such as around the shock or any shell or torus features in the wind, 
must be defined and 
sufficiently resolved in the new grid.  The criteria used to define this new grid will change based on the details of 
each problem being studied.  Generally the interior of the star, from which luminosity cannot escape, and the stellar 
wind, which is not finely structured, can be included at low resolution.  A hybrid scheme which combines linear and 
logarithmic binning was used for all subsequent spectra in this paper and is compared to simpler schemes in 
Figure \ref{fig:comp_bin}.  This scheme includes finely spaced linear bins between the radiation front and the $\tau$=20 
surface and more widely spaced log bins inside that radius and in the wind.  The $\tau$=20 surface is calculated using the 
Thomson scattering opacity, and defines the minimum inner bound of the region from which luminosity can escape.  The RAGE 
data inside each of these new radial bins is mass-averaged to ensure that very sharp features are included.  

\begin{figure}
\epsscale{1.17}
\plotone{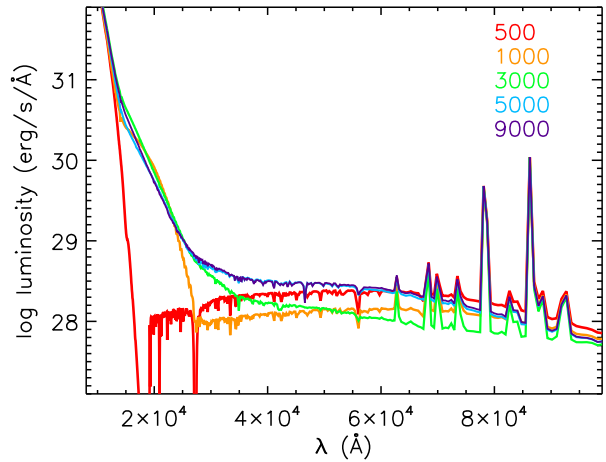}
\caption{
Radial convergence test, running SPECTRUM with the same RAGE data as \ref{fig:comp_bin} and 160 angular bins.  
This data is binned with varying radial resolution between the $\tau=20$ surface and the radiation front.  
Note that the spectra calculated with 5000 and 9000 radial bins are nearly identical.  
}
\vspace{0.1in}
\label{fig:radbin_zoom}
\end{figure}

To find the minimum number of bins that resolve the radiating region well, we performed a convergence test using between 
500 and 9,000 radial bins in the region between the $\tau$=20 surface and the radiation front, with the hybrid scheme 
described above.  For each of these tests, the interior region and the wind are each resolved by an additional 100 
logarithmically spaced bins.  As shown in Figure \ref{fig:radbin_zoom}, this SN Ic spectrum converges at about 5000 
radial bins.  The spectra in this figure were all calculated with 160 angular bins.  In addition to these SPECTRUM tests, 
we compared each binned parameter with the complete RAGE data set to confirm that the convergence shown here does 
correspond to a well resolved profile.  

\subsubsection{Angular Convergence}\label{sec:ang_bin}

\begin{figure}
\epsscale{1.17}
\plotone{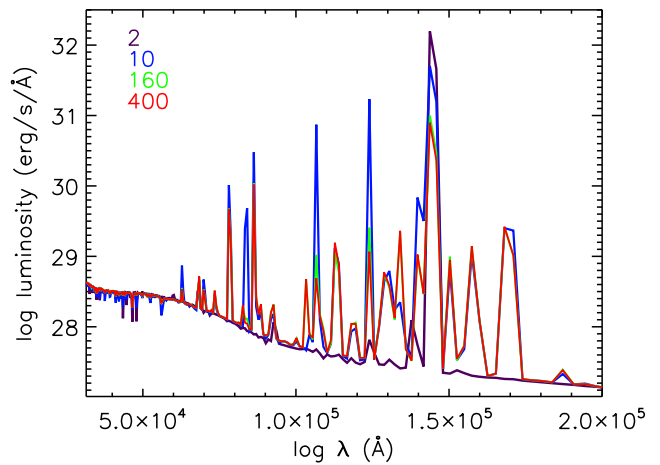}
\caption{
Angular convergence test, running SPECTRUM with the same RAGE data as \ref{fig:comp_bin}, 5200 radial bins and a varying 
number of angular bins.  
} \vspace{0.1in}
\label{fig:angbin_zoom}
\end{figure}

When the number of angular bins is increased in SPECTRUM, a finer grid is used to calculate luminosity along the 
line of sight for a single grid point (see Figure \ref{fig:grid1}.  This results in a more accurate luminosity 
calculation.  A convergence study was performed with between 2 and 400 angular bins and a fixed number of radial 
bins.  Figure \ref{fig:angbin_zoom} shows that any number of angular bins greater than two gives a nearly identical 
continuum and that line intensities converge at 160 angular bins, with 5200 radial bins.  However, we note that higher 
resolution might be required for detailed studies of individual lines.  We again emphasize that convergence in general 
is problem-dependent because the optimum scheme for allocating grid points depends on the structure of the radiating 
flow.  

\section{Physics Effects of Radiation Hydrodynamics} \label{sec:rage_tests}

The numerical calculations of SN light curves performed to date have implemented varying degrees of physics, 
as noted in Section \ref{sec:rage}.  To investigate the importance of radiation processes that have been neglected 
in past studies of SN explosions, we performed five 1D simulations of the Type Ic SN described in
Section \ref{sec:conv} in which additional physics was incrementally activated in RAGE.  Because our goal is
to examine how radiation physics affects SN profiles, spectra and light curves, we make no attempt to compare 
hydrodynamical algorithms.  The five cases we consider are as follows:

\begin{itemize}
\item{2T:  Gray flux-limited diffusion (FLD) in which matter and radiation temperatures are evolved separately 
in the simulation.  This physics option is used in all the simulations presented in subsequent sections.}
\item{1T:  Gray FLD in which matter and radiation are assumed to have the same temperature.} 
\item{low-$\kappa$: 2T gray FLD with $\kappa =$ 0.4 $\mathrm{cm}^2/ \mathrm{g}$ (the simple analytic scattering 
opacity for ionized H) for each constituent for all densities and temperatures.}  
\item{high-$\kappa$:  2T gray FLD with $\kappa = 10^4$ cm$^2/$g for all densities and temperatures.  
Many astrophysics calculations have been done with codes that include a radiation pressure proportional to $T^4$ 
in the EOS but do not perform radiation transport.  Setting the opacity to a large constant value in RAGE can 
serve as a proxy for radiation pressure in the EOS because energy will be contained in the radiation field but 
not transported, except in extremely low density regions.  This method allows gas temperatures to account for 
the presence of the radiation and prevents them from spiking to extremely high values in low density regions 
when internal energy is deposited in the cells (by shocks, for example).}
\item{hydro-only:  no radiation transport is performed, just hydrodynamics, and there are no radiation energies 
in the EOS. Because of its simplicity, the hydro-only option is useful for a wide variety of astrophysical problems 
in which radiation energies are negligible in comparison to internal energies.  However, in problems such as 
SNe where regions of the domain are dominated by radiation, this method can lead to highly unphysical results.}
\end{itemize}
These models were evolved in RAGE to well past shock breakout, which occurs at $\sim$ 0.7 s.

\begin{figure*}
\epsscale{0.65}
\plotone{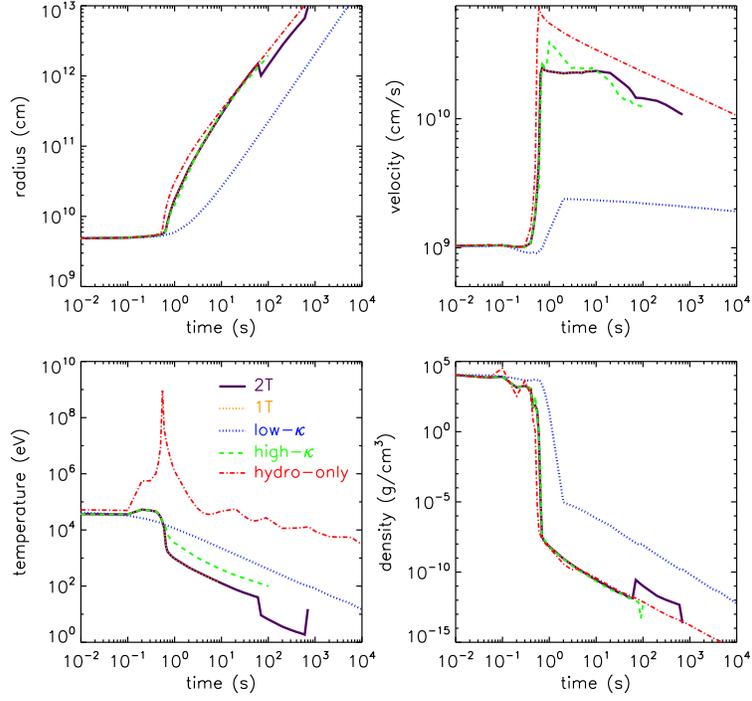}
\caption{Gas velocities, temperatures and densities in the cell with the maximum gas velocity on the grid, along
with the position of this cell, as a function of time for a Type Ic SN simulation in RAGE run with five different 
radiation/opacity treatments.  The data from the 1T simulation is indistinguishable from those of the 2T results.
}
\label{fig:rageshock}
\vspace{0.1in}
\end{figure*}

\begin{figure*}
\epsscale{0.7}
\plotone{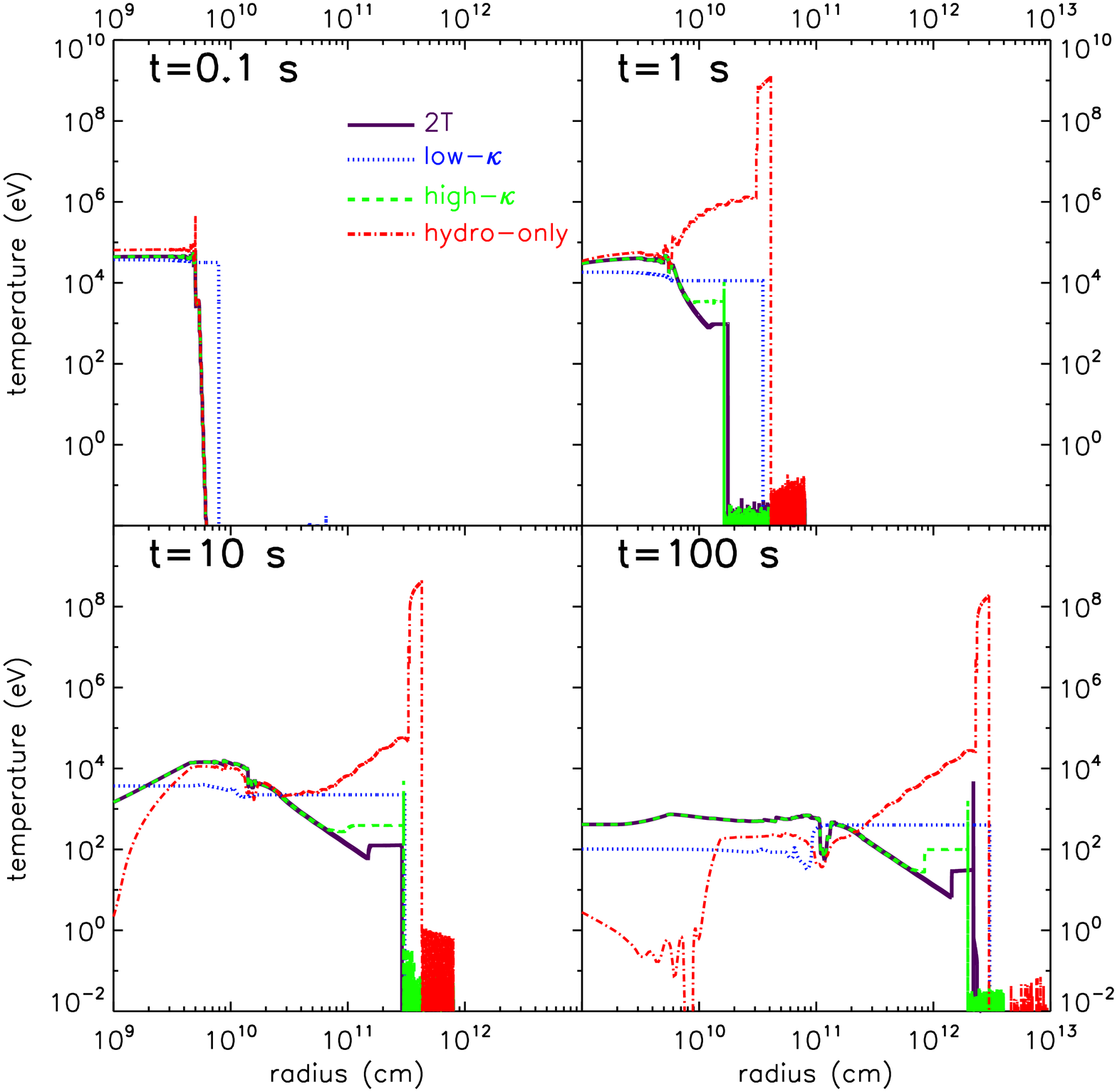}
\caption{Gas temperature profiles at 0.1, 1.0, 10.0, and 100.0 s for a SN Type Ic for our five tests.}
\label{fig:ragetev}
\vspace{0.1in}
\end{figure*}

\begin{figure*}
\epsscale{0.7}
\plotone{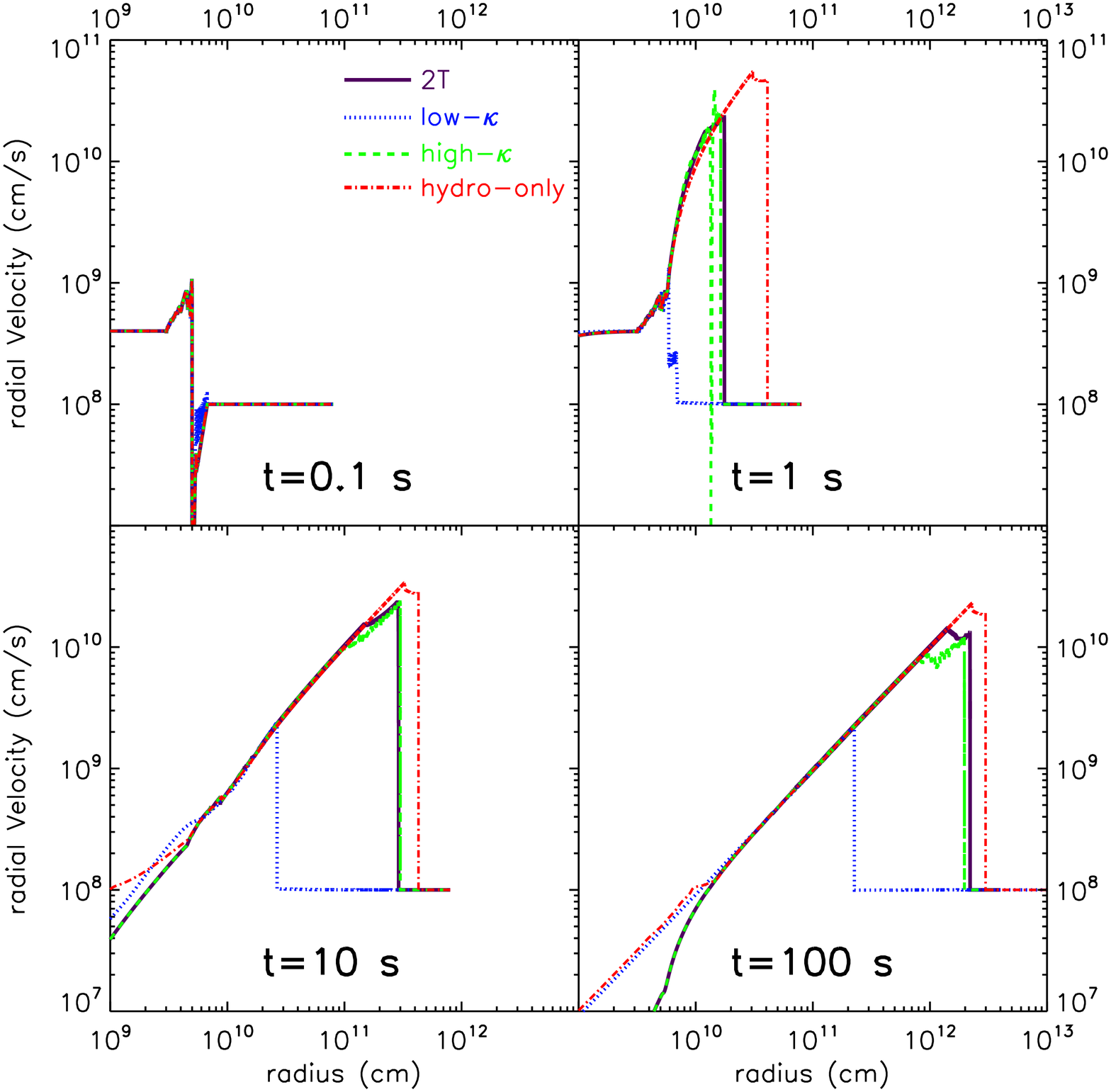}
\caption{Velocity profiles at 0.1, 1.0, 10.0, and 100.0 s in a SN Type Ic for our five tests.}
\label{fig:ragevel}
\vspace{0.1in}
\end{figure*}

\begin{figure*}
\epsscale{0.7}
\plotone{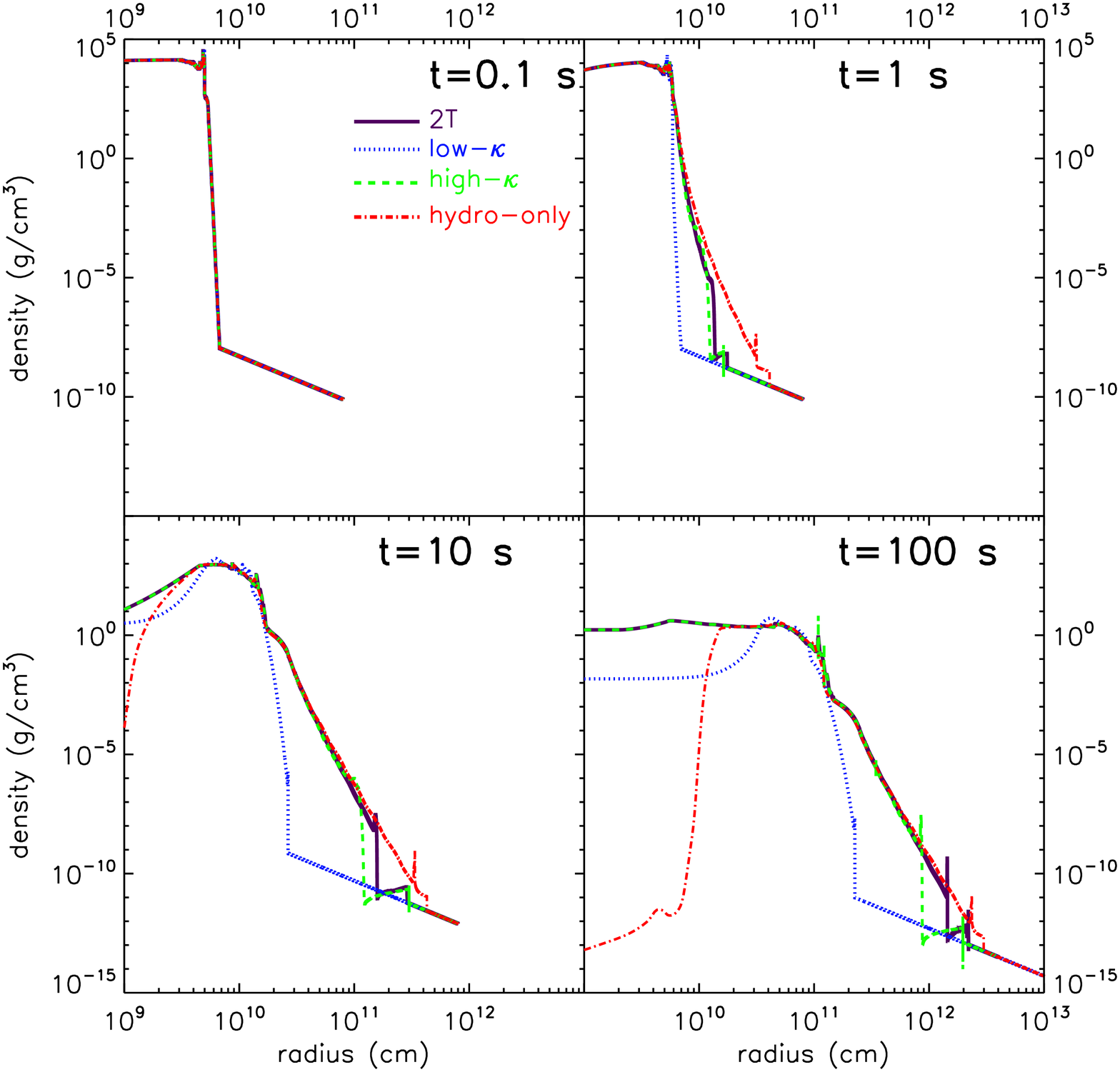}
\caption{Density profiles at 0.1, 1.0, 10.0, and 100.0 s for a SN Type Ic for our five tests.}
\label{fig:rageden}
\vspace{0.2in}
\end{figure*}

The interior of the SN evolves differently in the five simulations, but this region is extremely optically 
thick and has little effect on the luminosity of the SN.  The photosphere of the shock, where gas abruptly becomes 
optically thin, primarily determines the observational properties of the SN.  We plot gas velocities, temperatures 
and densities for the cell that has the maximum gas velocity on the grid, along with the position of this cell, as a 
function of time for all five tests in Figure \ref{fig:rageshock}.  

We note that the 1T and 2T simulation results are nearly identical throughout the duration of the 1T simulation 
because radiation is still trapped. After radiation breakout, the explicit solver used by the 1T simulation requires 
the use of extremely small time steps and it is impractical to run the simulation to completion.
The hydro-only simulation has the fastest shock, with velocities well above the 
speed of light.  Shock temperatures in this model approach 1 GeV at breakout.  These extreme velocities and 
temperatures arise because in a pure hydrodynamic simulation kinetic energy from the shock can only be 
converted into internal energy when the shock breaks out into the low-density wind.  In reality, this energy 
should be partitioned between the internal energy of the gas $E_{int}$ ($\propto T$) and the radiation energy 
$E_{rad}$ ($\propto T^4$).  In low density regions like the wind, most of the energy goes into radiation, and 
because the energy released by the shock is dominated by the $T^4$ term and the gas and radiation 
temperatures remain coupled, the true gas temperature will be much lower.  

In the high-$\kappa$ test, velocities and temperatures in the cell with the highest gas velocity are consistently 
lower than in the hydro-only simulation.  Although radiation is not efficiently transported in this model, it more 
properly divides energy between gas and radiation when the shock breaks out into the wind.  The 2T and 1T 
models are very similar to the high-$\kappa$ model prior to breakout. After breakout, the 2T and 1T solutions still 
have similar velocities, but lower temperatures than the high-$\kappa$ case because the wind ahead of the shock 
is optically thin and radiation efficiently transports energy out of the shock and cools it.  The sudden shift in the 
2T model around 50 seconds is an artifact due to the noisy velocity profile behind the shock.  The maximum 
velocity is initially near the front of the shock but a higher velocity region forms slightly behind the shock, 
causing the maximum velocity to suddenly shift to a deeper position.  In the low-$\kappa$ test, the cell with the 
highest velocity trails well behind those in the other simulations and remains at a higher density. In this run, the 
maximum velocity traces the expansion of the outer edge of the star because the low opacity prevents radiation 
from imparting momentum to the wind and accelerating it, as in the other models with radiation. 

\begin{deluxetable}{crrr}
\tabletypesize{\scriptsize}
\tablecaption{Bolometric Luminosity}
\tablehead{
\colhead{Time} & \colhead{2T} & \colhead{high-$\kappa$} & \colhead{low-$\kappa$}}
\startdata
0.1   & $2.03 \times 10^{34}$ & $2.02 \times 10^{34}$  & $9.72 \times 10^{34}$ \\
1.0   & $2.95 \times 10^{34}$ & $3.25 \times 10^{36}$  & $6.09 \times 10^{41}$ \\
10    & $3.76 \times 10^{38}$ & $5.68 \times 10^{39}$  & $7.38 \times 10^{42}$ \\
100   & $1.95 \times 10^{38}$ & $5.22 \times 10^{40}$  & $1.98 \times 10^{42}$ \\
& \vspace{-0.05in}&  & 
\enddata
\label{tab:bl}
\end{deluxetable}

\begin{deluxetable}{crrr}
\tabletypesize{\scriptsize}
\tablecaption{V Band Magnitudes}
\tablehead{
\colhead{Time} & \colhead{2T} & \colhead{high-$\kappa$} & \colhead{low-$\kappa$}}
\startdata
0.1   & 179.5  & 179.5 &  100.3 \\
1.0   &  27.3  & 26.3  &  179.5 \\
10    &   1.8  & 10.0  &   -2.7 \\
100   &  -5.9  & -6.7  &  -8.3  \\
& \vspace{-0.05in}&  & 
\enddata
\label{tab:vmags}
\end{deluxetable}

\begin{deluxetable}{crrr}
\tabletypesize{\scriptsize}
\tablecaption{B Band Magnitudes}
\tablehead{
\colhead{Time} & \colhead{2T} & \colhead{high-$\kappa$} & \colhead{low-$\kappa$}}
\startdata

0.1   & 181.3  & 181.3 &  75.4 \\
1.0   &  23.0  & 23.2  &  181.3 \\
10    &  -2.1  &  8.9  &  -2.9 \\
100   &  -6.5  & -7.9  &  -10.1 \\
& \vspace{-0.05in}&  & 
\enddata
\label{tab:bmags}
\end{deluxetable}

This latter point can be more clearly seen in the temperature, density, and velocity profiles for the explosion in 
Figures \ref{fig:ragetev}-\ref{fig:rageden}.  In the low-$\kappa$ simulation, the radiation front is visible as the 
nearly uniform temperature region extending from the surface of the star into the wind at times greater than 0.1 
s in Figure \ref{fig:ragetev}.  Comparing the positions of the radiation front and the maximum velocity in Figure 
\ref{fig:ragevel}, it is clear that the radiation has propagated well into the wind without depositing any momentum 
in it since its velocity is still the initial value of $10^8$ cm/s.  In the other four models, the position of the 
radiation front coincides with that of the shock that the radiation drives into the wind ahead of the expanding outer 
layers of the star.  As shown in Figure \ref{fig:rageden}, this shock creates a more gradual transition in density 
between the wind and the outer edge of the ejecta.  The blowoff of the outer layers of the star changes the depth from 
which photons escape the flow.  The behavior of the shock propagating into low density wind is examined in detail by 
\citet{colgate66} and \citet{mm99}.

\begin{figure*}
\epsscale{1.15}
\plottwo{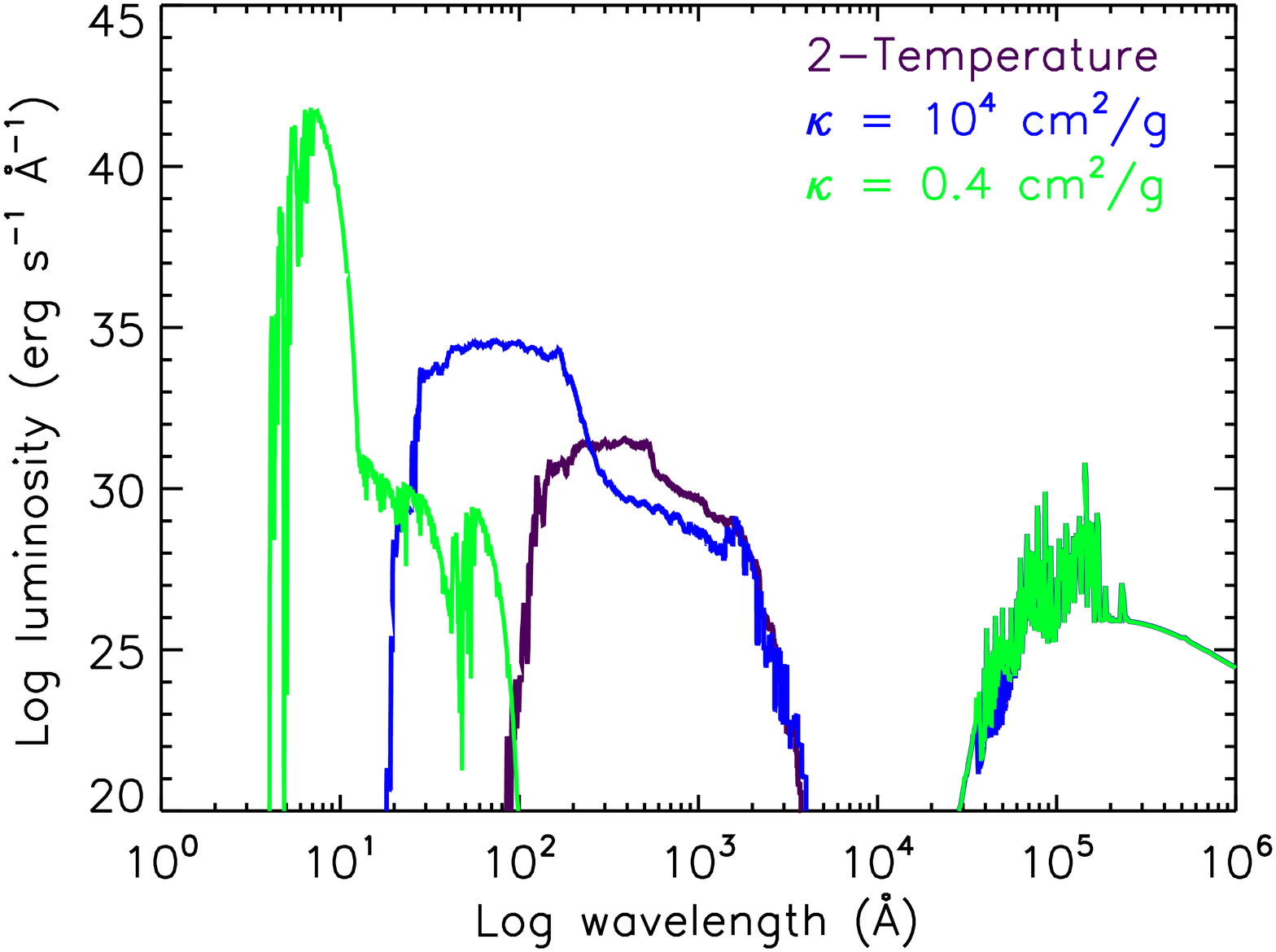}{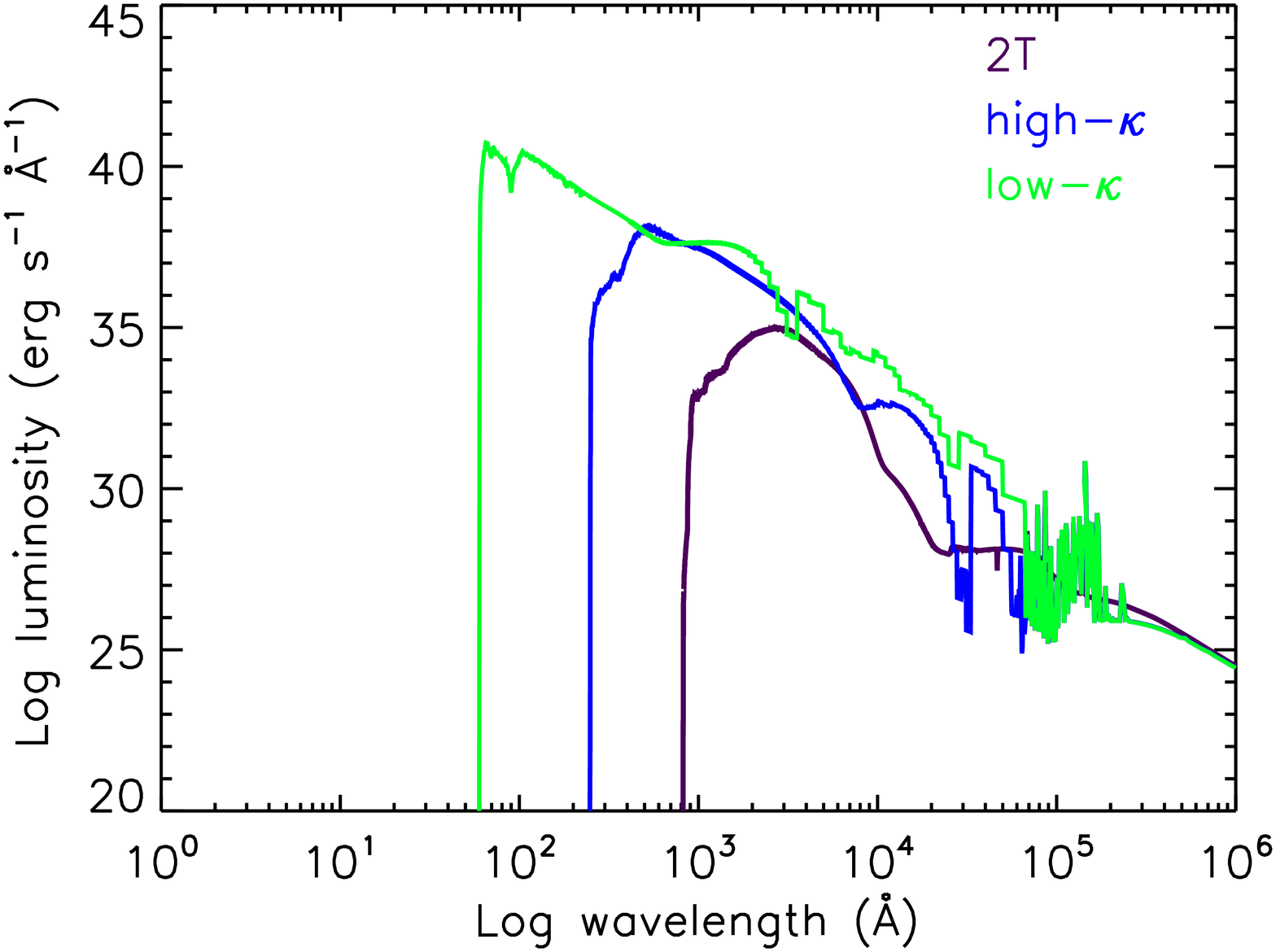}
\caption{Spectra for three different opacity treatment in RAGE.  Note the expansion and cooling of the fireball between 
1 second (left) and 100 seconds (right).  The 2T spectra are from the same simulation presented in Figures \ref{fig:tau1_rad}, 
\ref{fig:tau1_lum}, \ref{fig:lc} and \ref{fig:lumpos}. }
\label{fig:ragespec}
\vspace{0.1in}
\end{figure*}

We show spectra for the 2T, high-$\kappa$, and low-$\kappa$ models at 1 and 100 s in Figure 
\ref{fig:ragespec}.  They are calculated with SPECTRUM with the full set of monochromatic opacities described 
in Section \ref{sec:spectrum_opac}.  We exclude the hydro-only model because its 1 GeV temperatures yield 
spectra that are unphysical.  The spectra follow the same trends as in Figure \ref{fig:rageshock}: the emitting 
region is hottest in the low-$\kappa$ simulations, followed by the high-$\kappa$ and 2T simulations.  Bolometric 
luminosities 
for all three runs are shown in Table \ref{tab:bl}.  They vary by up to seven orders of magnitude just because of 
differences in opacities in the radiation hydrodynamics (RHD) calculations.  Their spectral profiles are also quite 
different, particularly at breakout.  We list V-band and B-band magnitudes for all three models in Tables \ref{tab:vmags} 
and \ref{tab:bmags}.  Depending on the band and the time, any of the three simulations can be brightest, often at 
variance with their relative bolometric 
luminosities.  At early times when they are below the limits of detection, luminosities in both bands can vary by 
more than 100 magnitudes with model.  At later times when they are above detection limits, luminosities in both 
bands can still vary by several magnitudes with model.  These extreme variations in bolometric luminosity and 
visible magnitudes demonstrate the importance of accurate opacities and RHD to realistic SN light curves and 
spectra.  Simulations with approximate opacities or no radiation transport can yield spectra with errors that 
are far greater than those in the observations.  This test case demonstrates a subset of all the possible effects of 
not modeling the full physics of a SN.  

\section{Physics Effects on Spectra} \label{sec:discussion}
\begin{figure}
\epsscale{1.17}
\plotone{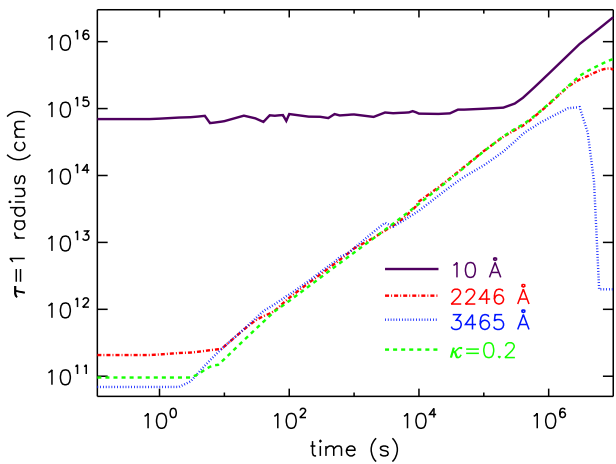}
\caption{
The radius of the $\tau$=1 surface vs. time, calculated at single wavelengths in 
the x-ray (10 \AA), m2 (2246 \AA) and u (3465 \AA) bands, and with a constant helium opacity ($\kappa=0.2$).  
}
\vspace{0.1in}
\label{fig:tau1_rad}
\end{figure}

\begin{figure}
\epsscale{1.17}
\plotone{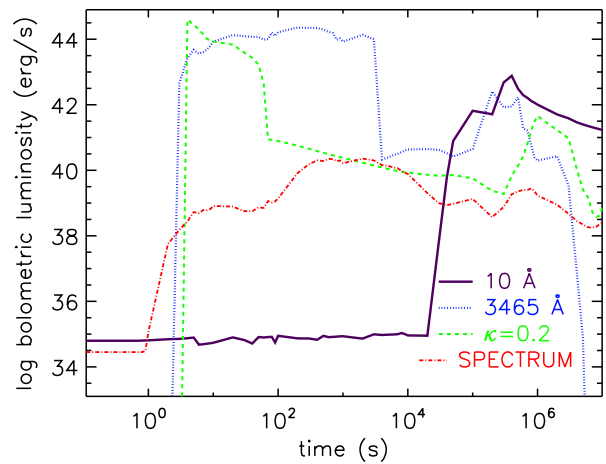}
\caption{
Bolometric light curves calculated from the Stefan-Boltzmann law at $\tau$=1 radii
determined with monochromatic opacities in the x-ray (10 \AA) and u (3465 \AA)  bands 
and a constant helium opacity ($\kappa=0.2$), overlaid with the bolometric light curve from SPECTRUM.
}
\vspace{0.1in}
\label{fig:tau1_lum}
\end{figure}

\begin{figure}
\epsscale{1.17}
\plotone{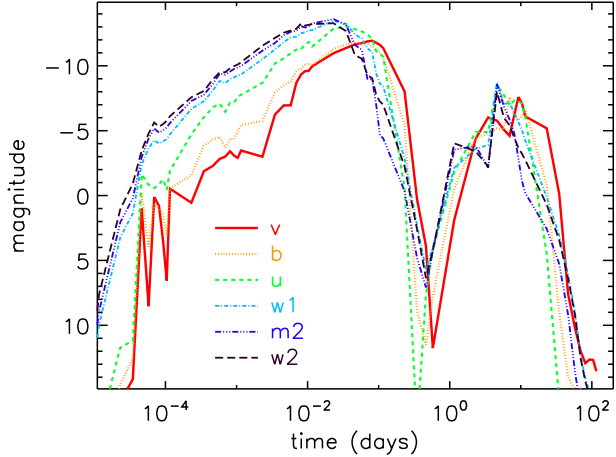}
\caption{
SPECTRUM light curves for six Swift UV/optical bands: v (5468 \AA), b (4392 \AA), u (3465 \AA), w1 (2600 \AA), 
m2 (2246 \AA), and w2 (1928 \AA).
}
\vspace{0.1in}
\label{fig:lc}
\end{figure}

\begin{figure}
\epsscale{1.17}
\plotone{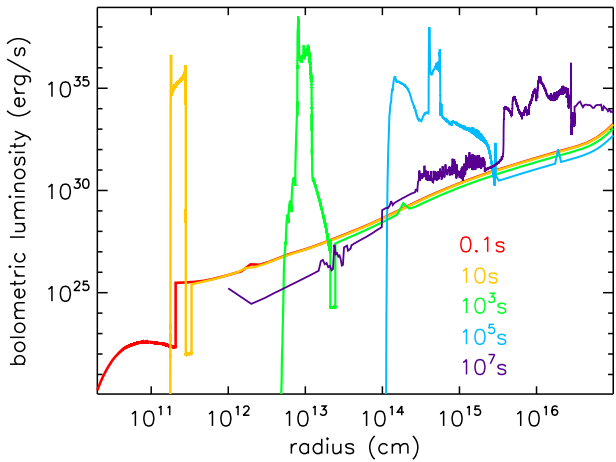}
\caption{
Luminosity as a function of emission radius calculated with SPECTRUM, for five different times. 
}
\vspace{0.1in}
\label{fig:lumpos}
\end{figure}

Analyses of shock breakout observations generally assume that the opacity is constant when deriving the radius at 
which radiation breaks free of the shock (i.e. \citealt{sod08}).  In reality, radiation does not break free of the 
shock or wind at a single radius or time because the opacities of the outer layers of the star, ejecta or surrounding 
wind are all strongly dependent on wavelength, temperature and density.  As shown in Figure 
\ref{fig:H_opac}, monochromatic opacities can vary by many orders of magnitude with wavelength, so a single radius and 
temperature cannot accurately describe this phenomenon.  Figures \ref{fig:tau1_rad}, \ref{fig:tau1_lum}, \ref{fig:lc} 
and \ref{fig:lumpos} all present data from the Type Ic SN progenitor described in Section \ref{sec:conv}.  

To study radiation breakout, we calculated the $\tau$=1 surface with monochromatic opacities for several 
wavelengths as well as a constant ionized helium opacity, $\kappa$=0.2.  Figure \ref{fig:tau1_rad} shows that for the 
wind of this SN opacity is high at short wavelengths, leading to a large $\tau$=1 radius.  After the shock breaks through 
the surface of the star and 
reaches the $\tau =$ 1 surface for a given wavelength, the surface coincides with that of the shock thereafter and 
expands with it.  As the ejecta expands it becomes spherically diluted and cools, causing the photosphere to retreat 
into the ejecta at longer wavelengths at later times, as seen in the u-band after $\sim$ 10$^6$ s.  

Because radiation breaks free of the shock at a range of radii, luminosity estimates can vary widely with the choice 
of $\tau =$ 1 surface, as we show in Figure \ref{fig:tau1_lum}.  Here we plot bolometric luminosities derived from the 
Stefan-Boltzmann law, with the radius and temperature at the $\tau =$ 1 surfaces shown in Figure \ref{fig:tau1_rad}.  
We contrast these semi-analytic light-curves with the bolometric luminosity derived from SPECTRUM calculations, 
which include luminosity 
emitted at multiple wavelengths and radii.  It is clear that the bolometric light curves peak at very different 
times depending on the choice of photosphere, and that the single-radius calculations dramatically overestimate or 
underestimate the luminosity.  In the $\kappa$=0.2 and u-band cases, the overestimates are partly because the Stefan-Boltzmann law 
fails to include flux attenuation by the envelope.  The wind, primarily consisting of oxygen, has high opacities at X-ray 
wavelengths, which results in the large $\tau=$1 radii in Figure \ref{fig:tau1_rad}.  The temperature at these 
surfaces is low until the shock has propagated sufficiently far for the radiative precursor, and then the shock, to 
raise the temperature.  This results in a bolometric luminosity from the X-ray $\tau$=1 surface that is initially 
constant and then abruptly rises at $\sim 2\times 10^{4}$ s, as the radiative precursor and then the shock heat the 
region outside the $\tau=$1 surface and increase the opacity.  The late-time (after $\sim 10^4$ s) bumps in the $\kappa$=0.2 
and u-band light curves occur as the shock propagates outward and the $\tau=$1 surface for the given 
wavelength descends into hotter, more dense regions of the shock.  The effect of radiation breakout occurring at different 
times and radii for various wavelengths is further illustrated in Figure \ref{fig:lc}, where we show light curves for six 
\textit{Swift} ultraviolet and optical bands calculated with SPECTRUM.   

Some of the differences between the semi-analytic light curves and the SPECTRUM light curve in Figure 
\ref{fig:tau1_lum} are due to emission and absorption in the flow on either side of the shock, which cannot be taken 
into account by a simple Stefan-Boltzmann calculation.  SPECTRUM calculates the luminosity that escapes the ejecta as 
a function of both wavelength and radius, enabling us to study luminosity as function of both radius and time as shown 
in Figure \ref{fig:lumpos}.  At 10, $10^3$ and $10^5$ s the double-peaked structure of the radiating shock is clearly 
visible.  This feature is absent at 0.1 s because the shock is still inside the star, and is smeared out at $10^7$ s 
because the ejecta has greatly expanded and become nearly transparent.  Shortly after breakout, which occurs at 
$\sim 0.7$ s, luminosity primarily originates from the relatively narrow shocked region itself but over time a 
radiative precursor propagates ahead of it, heats the material there, and causes it to radiate as well.  As the ejecta 
expands and is spherically diluted, radiation can escape from deeper in its interior.  This effect is visible in the 
profiles as a plateau that begins to extend from the rear surface of the radiating shock at $10^3$ s.  

\begin{figure}
\epsscale{1.17}
\plotone{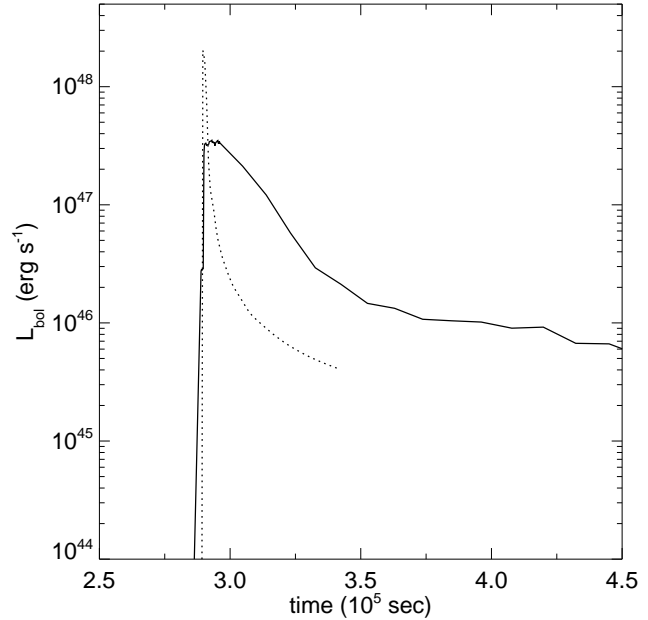}
\caption{Bolometric lightcurves, calculated analytically (dotted line) and with SPECTRUM (solid line), for the 
pair-instability supernova of a 225 M$_{\odot}$ blue compact Pop III star.}
\label{fig:breakout}
\vspace{0.05in}
\end{figure}

If all the photons broke free of the shock at once, the duration of the initial transient would be comparable to the 
light-crossing time of the star, since photons emitted from its poles and its equator would reach an observer at 
times that differ by the time it takes light to cross the star. However, radiation-matter coupling generally smears 
radiation breakout over much longer times than the light-crossing time of the star, as shown in  
Figure \ref{fig:breakout}.  Here, we plot bolometric luminosities at shock breakout for a 225 M$_{\odot}$ Pop III 
pair-instability supernova evolved in KEPLER and RAGE.  Strong coupling between photons and matter causes the 
radiation front to blow off the outermost layers of the star with very high velocities, but this in turn impedes the 
escape of the photons into the surrounding medium.  This delay is evident in the relative widths of the bolometric 
luminosities calculated by SPECTRUM and with the Stefan-Boltzmann law (we use $\kappa_{Th} =$ 0.288, the opacity due 
to Thomson scattering in primordial H and He gas, to compute the radius of the $\tau =$ 1 surface).  The width of the 
breakout transient calculated analytically is roughly the light-crossing time of the star, given that its radius is a 
few $\times$ 10$^{13}$ cm, but the breakout pulse calculated with RAGE and SPECTRUM is both broader and dimmer because 
photons remain coupled to the shock for longer times, as seen in Figures \ref{fig:tau1_lum} and \ref{fig:breakout}. 

\section{Conclusion} \label{sec:conclusion}

Current and future all-sky surveys will provide numerous light curves from supernovae and other transients, including 
many which do not fit a previously known type (for example, 3$\%$ of PTF SNe are unclassified as of March 
2012\footnote{www.astro.caltech.edu/ptf/}).  However, light curves only provide indirect measurements of the physical 
properties of progenitor stars and their surrounding environments.  
Analytical calculations performed to derive these properties assume a constant opacity and treat shock breakout as a single event.  
Approximations made in some hydro-only or RHD codes result in inaccurate opacities or incompletely 
coupled radiation and hydrodynamics, as demonstrated in Section \ref{sec:rage_tests}.  
To correct some of the 
inaccuracies inherent in these approaches, we have developed a new pipeline starting with RHD 
simulations in RAGE including accurate gray opacities.  We then post-process these simulations with SPECTRUM, 
computing emitted and absorbed luminosity from accurate monochromatic opacities.  This pipeline allows us to 
accurately model the luminosity from SNe as a function of radius, wavelength and time.  

Our RAGE and SPECTRUM tests show that shock breakout and radiation breakout in supernova explosions 
are inherently radiation hydrodynamical and wavelength dependent in nature, and cannot be accurately modeled by 
current analytical methods or numerical simulations that lack this physics.  Doing so can lead to serious errors when 
deriving the properties of supernovae, and hence their progenitors and environments, from light curve observations.  
We also find that inaccurate opacities can result in significant errors in light curves and spectra, even well after 
shock breakout.  Another key to calculating realistic SN light curves and spectra is ensuring that both simulations 
and postprocessing resolve the photosphere of the shock at all times, since this region of the flow is most important 
to its observational properties.  

The sensitivity of our tests to opacities and radiation physics and 
the detailed structure seen in spectra from monochromatic opacities suggest that even more accurate spectra can be 
obtained with multi-group radiation transport.  However, because multi-group 
simulation times roughly scale as the number of groups, it is computationally prohibitive to run with a large number 
of groups.  The results from a single-group RAGE simulation post-processed with SPECTRUM can direct the choice of 
energy groups to cover the range of expected energies and have appropriate resolution near dominant spectral features.

The pipeline we have developed with RAGE and SPECTRUM is now being used to model the observational signatures of a 
broad range of SNe. Core-collapse supernovae in dense winds are being studied in which radiation breakout occurs 
in the wind, well after shock breakout from the surface of the star (L.H. Frey et al., in preparation). We are also now investigating 
supernovae with more complex environments, such as Type Ia explosions with circumstellar tori (W. Even et al., in preparation) and Pop 
III core-collapse SNe in dense circumstellar shells (Type IIn SNe, whose brightness can rival those of much more 
massive pair-instability SNe, D.J. Whalen et al., in preparation). We have calculated light curves and spectra for both Pop III core-collapse 
and pair-instability SNe in order to determine out to what redshifts they can be detected by \textit{JWST} and future 
thirty-meter telescopes \citep{wet12c,wet12b, wet12a}, and are also modeling the explosions of supermassive Pop III 
stars in $z \sim$ 15 protogalaxies, and pair-instability SNe in the local universe (C.C. Joggerst et al., in preparation).  The 
flexibility of our approach will yield many insights into these astrophysical transients as well as future ones 
whose progenitors are yet to be characterized.

\acknowledgments

We thank Gabriel Rockefeller of CCS-2 at LANL for the development of the code used to extract explosion 
profiles from RAGE AMR data.  DJW  was supported by the Bruce and Astrid McWilliams Center for Cosmology 
at Carnegie Mellon University.  Work at LANL was done under the auspices of the National Nuclear Security 
Administration of the U.S. Department of Energy at Los Alamos National Laboratory under Contract No. 
DE-AC52-06NA25396.  All RAGE and SPECTRUM calculations were performed on Institutional Computing (IC) 
and Yellow network platforms at LANL (Conejo, Mapache, Lobo, Turing and Yellowrail).

\bibliographystyle{apj}
\bibliography{refs}

\begin{thebibliography}{60}
\expandafter\ifx\csname natexlab\endcsname\relax\def\natexlab#1{#1}\fi

\bibitem[{{Almgren} {et~al.}(2010){Almgren}, {Beckner}, {Bell}, {Day},
  {Howell}, {Joggerst}, {Lijewski}, {Nonaka}, {Singer}, \&
  {Zingale}}]{Almgren2010}
{Almgren}, A.~S., {Beckner}, V.~E., {Bell}, J.~B., {Day}, M.~S., {Howell},
  L.~H., {Joggerst}, C.~C., {Lijewski}, M.~J., {Nonaka}, A., {Singer}, M., \&
  {Zingale}, M. 2010, \apj, 715, 1221

\bibitem[{{Balberg} \& {Loeb}(2011)}]{bl11}
{Balberg}, S. \& {Loeb}, A. 2011, \mnras, 414, 1715

\bibitem[{{Blinnikov} {et~al.}(2000){Blinnikov}, {Lundqvist}, {Bartunov},
  {Nomoto}, \& {Iwamoto}}]{Blinn00}
{Blinnikov}, S., {Lundqvist}, P., {Bartunov}, O., {Nomoto}, K., \& {Iwamoto},
  K. 2000, \apj, 532, 1132

\bibitem[{{Blinnikov} {et~al.}(1998){Blinnikov}, {Eastman}, {Bartunov},
  {Popolitov}, \& {Woosley}}]{Blinn98}
{Blinnikov}, S.~I., {Eastman}, R., {Bartunov}, O.~S., {Popolitov}, V.~A., \&
  {Woosley}, S.~E. 1998, \apj, 496, 454

\bibitem[{{Branch} {et~al.}(1985){Branch}, {Doggett}, {Nomoto}, \&
  {Thielemann}}]{Branch85}
{Branch}, D., {Doggett}, J.~B., {Nomoto}, K., \& {Thielemann}, F.-K. 1985,
  \apj, 294, 619

\bibitem[{{Campana} {et~al.}(2006){Campana}, {Mangano}, {Blustin}, {Brown},
  {Burrows}, {Chincarini}, {Cummings}, {Cusumano}, {Della Valle}, {Malesani},
  {M{\'e}sz{\'a}ros}, {Nousek}, {Page}, {Sakamoto}, {Waxman}, {Zhang}, {Dai},
  {Gehrels}, {Immler}, {Marshall}, {Mason}, {Moretti}, {O'Brien}, {Osborne},
  {Page}, {Romano}, {Roming}, {Tagliaferri}, {Cominsky}, {Giommi}, {Godet},
  {Kennea}, {Krimm}, {Angelini}, {Barthelmy}, {Boyd}, {Palmer}, {Wells}, \&
  {White}}]{camp06}
{Campana}, S., {Mangano}, V., {Blustin}, A.~J., {Brown}, P., {Burrows}, D.~N.,
  {Chincarini}, G., {Cummings}, J.~R., {Cusumano}, G., {Della Valle}, M.,
  {Malesani}, D., {M{\'e}sz{\'a}ros}, P., {Nousek}, J.~A., {Page}, M.,
  {Sakamoto}, T., {Waxman}, E., {Zhang}, B., {Dai}, Z.~G., {Gehrels}, N.,
  {Immler}, S., {Marshall}, F.~E., {Mason}, K.~O., {Moretti}, A., {O'Brien},
  P.~T., {Osborne}, J.~P., {Page}, K.~L., {Romano}, P., {Roming}, P.~W.~A.,
  {Tagliaferri}, G., {Cominsky}, L.~R., {Giommi}, P., {Godet}, O., {Kennea},
  J.~A., {Krimm}, H., {Angelini}, L., {Barthelmy}, S.~D., {Boyd}, P.~T.,
  {Palmer}, D.~M., {Wells}, A.~A., \& {White}, N.~E. 2006, \nat, 442, 1008

\bibitem[{{Chevalier} \& {Fransson}(2008)}]{chev08}
{Chevalier}, R.~A. \& {Fransson}, C. 2008, \apjl, 683, L135

\bibitem[{{Chevalier} \& {Irwin}(2011)}]{chev11}
{Chevalier}, R.~A. \& {Irwin}, C.~M. 2011, \apjl, 729, L6+

\bibitem[{{Colgate}(1974)}]{cg74}
{Colgate}, S.~A. 1974, \apj, 187, 333

\bibitem[{{Colgate} \& {White}(1966)}]{colgate66}
{Colgate}, S.~A. \& {White}, R.~H. 1966, \apj, 143, 626

\bibitem[{{Dwek} \& {Arendt}(2008)}]{dwek08}
{Dwek}, E. \& {Arendt}, R.~G. 2008, \apj, 685, 976

\bibitem[{{Eastman} \& {Pinto}(1993)}]{EP93}
{Eastman}, R.~G. \& {Pinto}, P.~A. 1993, \apj, 412, 731

\bibitem[{{Ensman} \& {Burrows}(1992)}]{Ens92}
{Ensman}, L. \& {Burrows}, A. 1992, \apj, 393, 742

\bibitem[{{Fryer}(1999)}]{fryer99}
{Fryer}, C.~L. 1999, \apj, 522, 413

\bibitem[{{Fryer}(2009)}]{fryer09}
---. 2009, \apj, 699, 409

\bibitem[{{Fryer} {et~al.}(2009){Fryer}, {Brown}, {Bufano}, {Dahl}, {Fontes},
  {Frey}, {Holland}, {Hungerford}, {Immler}, {Mazzali}, {Milne}, {Scannapieco},
  {Weinberg}, \& {Young}}]{fet09}
{Fryer}, C.~L., {Brown}, P.~J., {Bufano}, F., {Dahl}, J.~A., {Fontes}, C.~J.,
  {Frey}, L.~H., {Holland}, S.~T., {Hungerford}, A.~L., {Immler}, S.,
  {Mazzali}, P., {Milne}, P.~A., {Scannapieco}, E., {Weinberg}, N., \& {Young},
  P.~A. 2009, \apj, 707, 193

\bibitem[{{Fryer} {et~al.}(1999){Fryer}, {Colgate}, \& {Pinto}}]{fcp99}
{Fryer}, C.~L., {Colgate}, S.~A., \& {Pinto}, P.~A. 1999, \apj, 511, 885

\bibitem[{{Fryer} {et~al.}(2006{\natexlab{a}}){Fryer}, {Herwig}, {Hungerford},
  \& {Timmes}}]{fhh06}
{Fryer}, C.~L., {Herwig}, F., {Hungerford}, A., \& {Timmes}, F.~X.
  2006{\natexlab{a}}, \apjl, 646, L131

\bibitem[{{Fryer} {et~al.}(2006{\natexlab{b}}){Fryer}, {Rockefeller}, \&
  {Warren}}]{frw06}
{Fryer}, C.~L., {Rockefeller}, G., \& {Warren}, M.~S. 2006{\natexlab{b}}, \apj,
  643, 292

\bibitem[{{Fryer} {et~al.}(2010){Fryer}, {Ruiter}, {Belczynski}, {Brown},
  {Bufano}, {Diehl}, {Fontes}, {Frey}, {Holland}, {Hungerford}, {Immler},
  {Mazzali}, {Meakin}, {Milne}, {Raskin}, \& {Timmes}}]{Fryer10}
{Fryer}, C.~L., {Ruiter}, A.~J., {Belczynski}, K., {Brown}, P.~J., {Bufano},
  F., {Diehl}, S., {Fontes}, C.~J., {Frey}, L.~H., {Holland}, S.~T.,
  {Hungerford}, A.~L., {Immler}, S., {Mazzali}, P., {Meakin}, C., {Milne},
  P.~A., {Raskin}, C., \& {Timmes}, F.~X. 2010, \apj, 725, 296

\bibitem[{{Gardner} {et~al.}(2006){Gardner}, {Mather}, {Clampin}, {Doyon},
  {Greenhouse}, {Hammel}, {Hutchings}, {Jakobsen}, {Lilly}, {Long}, {Lunine},
  {McCaughrean}, {Mountain}, {Nella}, {Rieke}, {Rieke}, {Rix}, {Smith},
  {Sonneborn}, {Stiavelli}, {Stockman}, {Windhorst}, \& {Wright}}]{JWST}
{Gardner}, J.~P., {Mather}, J.~C., {Clampin}, M., {Doyon}, R., {Greenhouse},
  M.~A., {Hammel}, H.~B., {Hutchings}, J.~B., {Jakobsen}, P., {Lilly}, S.~J.,
  {Long}, K.~S., {Lunine}, J.~I., {McCaughrean}, M.~J., {Mountain}, M.,
  {Nella}, J., {Rieke}, G.~H., {Rieke}, M.~J., {Rix}, H.-W., {Smith}, E.~P.,
  {Sonneborn}, G., {Stiavelli}, M., {Stockman}, H.~S., {Windhorst}, R.~A., \&
  {Wright}, G.~S. 2006, \ssr, 123, 485

\bibitem[{{Gehmeyr} \& {Mihalas}(1994)}]{Gehmeyr94}
{Gehmeyr}, M. \& {Mihalas}, D. 1994, Physica D Nonlinear Phenomena, 77, 320

\bibitem[{{Gezari} {et~al.}(2008){Gezari}, {Dessart}, {Basa}, {Martin},
  {Neill}, {Woosley}, {Hillier}, {Bazin}, {Forster}, {Friedman}, {Le Du},
  {Mazure}, {Morrissey}, {Neff}, {Schiminovich}, \& {Wyder}}]{gez08}
{Gezari}, S., {Dessart}, L., {Basa}, S., {Martin}, D.~C., {Neill}, J.~D.,
  {Woosley}, S.~E., {Hillier}, D.~J., {Bazin}, G., {Forster}, K., {Friedman},
  P.~G., {Le Du}, J., {Mazure}, A., {Morrissey}, P., {Neff}, S.~G.,
  {Schiminovich}, D., \& {Wyder}, T.~K. 2008, \apjl, 683, L131

\bibitem[{{Gittings} {et~al.}(2008){Gittings}, {Weaver}, {Clover}, {Betlach},
  {Byrne}, {Coker}, {Dendy}, {Hueckstaedt}, {New}, {Oakes}, {Ranta}, \&
  {Stefan}}]{rage}
{Gittings}, M., {Weaver}, R., {Clover}, M., {Betlach}, T., {Byrne}, N.,
  {Coker}, R., {Dendy}, E., {Hueckstaedt}, R., {New}, K., {Oakes}, W.~R.,
  {Ranta}, D., \& {Stefan}, R. 2008, Computational Science and Discovery, 1,
  015005

\bibitem[{{Hauschildt} \& {Ensman}(1994)}]{HE94}
{Hauschildt}, P.~H. \& {Ensman}, L.~M. 1994, \apj, 424, 905

\bibitem[{{Hoeflich} {et~al.}(1993){Hoeflich}, {Mueller}, \&
  {Khokhlov}}]{Hoeflich93}
{Hoeflich}, P., {Mueller}, E., \& {Khokhlov}, A. 1993, \aap, 268, 570

\bibitem[{{Ivezic} {et~al.}(2008){Ivezic}, {Tyson}, {Acosta}, {Allsman},
  {Anderson}, {Andrew}, {Angel}, {Axelrod}, {Barr}, {Becker}, {Becla},
  {Beldica}, {Blandford}, {Bloom}, {Borne}, {Brandt}, {Brown}, {Bullock},
  {Burke}, {Chandrasekharan}, {Chesley}, {Claver}, {Connolly}, {Cook},
  {Cooray}, {Covey}, {Cribbs}, {Cutri}, {Daues}, {Delgado}, {Ferguson},
  {Gawiser}, {Geary}, {Gee}, {Geha}, {Gibson}, {Gilmore}, {Gressler}, {Hogan},
  {Huffer}, {Jacoby}, {Jain}, {Jernigan}, {Jones}, {Juric}, {Kahn}, {Kalirai},
  {Kantor}, {Kessler}, {Kirkby}, {Knox}, {Krabbendam}, {Krughoff}, {Kulkarni},
  {Lambert}, {Levine}, {Liang}, {Lim}, {Lupton}, {Marshall}, {Marshall}, {May},
  {Miller}, {Mills}, {Monet}, {Neill}, {Nordby}, {O'Connor}, {Oliver},
  {Olivier}, {Olsen}, {Owen}, {Peterson}, {Petry}, {Pierfederici},
  {Pietrowicz}, {Pike}, {Pinto}, {Plante}, {Radeka}, {Rasmussen}, {Ridgway},
  {Rosing}, {Saha}, {Schalk}, {Schindler}, {Schneider}, {Schumacher}, {Sebag},
  {Seppala}, {Shipsey}, {Silvestri}, {Smith}, {Smith}, {Strauss}, {Stubbs},
  {Sweeney}, {Szalay}, {Thaler}, {Vanden Berk}, {Walkowicz}, {Warner},
  {Willman}, {Wittman}, {Wolff}, {Wood-Vasey}, {Yoachim}, {Zhan}, \& {for the
  LSST Collaboration}}]{LSST}
{Ivezic}, Z., {Tyson}, J.~A., {Acosta}, E., {Allsman}, R., {Anderson}, S.~F.,
  {Andrew}, J., {Angel}, R., {Axelrod}, T., {Barr}, J.~D., {Becker}, A.~C.,
  {Becla}, J., {Beldica}, C., {Blandford}, R.~D., {Bloom}, J.~S., {Borne}, K.,
  {Brandt}, W.~N., {Brown}, M.~E., {Bullock}, J.~S., {Burke}, D.~L.,
  {Chandrasekharan}, S., {Chesley}, S., {Claver}, C.~F., {Connolly}, A.,
  {Cook}, K.~H., {Cooray}, A., {Covey}, K.~R., {Cribbs}, C., {Cutri}, R.,
  {Daues}, G., {Delgado}, F., {Ferguson}, H., {Gawiser}, E., {Geary}, J.~C.,
  {Gee}, P., {Geha}, M., {Gibson}, R.~R., {Gilmore}, D.~K., {Gressler}, W.~J.,
  {Hogan}, C., {Huffer}, M.~E., {Jacoby}, S.~H., {Jain}, B., {Jernigan}, J.~G.,
  {Jones}, R.~L., {Juric}, M., {Kahn}, S.~M., {Kalirai}, J.~S., {Kantor},
  J.~P., {Kessler}, R., {Kirkby}, D., {Knox}, L., {Krabbendam}, V.~L.,
  {Krughoff}, S., {Kulkarni}, S., {Lambert}, R., {Levine}, D., {Liang}, M.,
  {Lim}, K., {Lupton}, R.~H., {Marshall}, P., {Marshall}, S., {May}, M.,
  {Miller}, M., {Mills}, D.~J., {Monet}, D.~G., {Neill}, D.~R., {Nordby}, M.,
  {O'Connor}, P., {Oliver}, J., {Olivier}, S.~S., {Olsen}, K., {Owen}, R.~E.,
  {Peterson}, J.~R., {Petry}, C.~E., {Pierfederici}, F., {Pietrowicz}, S.,
  {Pike}, R., {Pinto}, P.~A., {Plante}, R., {Radeka}, V., {Rasmussen}, A.,
  {Ridgway}, S.~T., {Rosing}, W., {Saha}, A., {Schalk}, T.~L., {Schindler},
  R.~H., {Schneider}, D.~P., {Schumacher}, G., {Sebag}, J., {Seppala}, L.~G.,
  {Shipsey}, I., {Silvestri}, N., {Smith}, J.~A., {Smith}, R.~C., {Strauss},
  M.~A., {Stubbs}, C.~W., {Sweeney}, D., {Szalay}, A., {Thaler}, J.~J., {Vanden
  Berk}, D., {Walkowicz}, L., {Warner}, M., {Willman}, B., {Wittman}, D.,
  {Wolff}, S.~C., {Wood-Vasey}, W.~M., {Yoachim}, P., {Zhan}, H., \& {for the
  LSST Collaboration}. 2008, arXiv: 0805.2366

\bibitem[{{Kaiser} {et~al.}(2002){Kaiser}, {Aussel}, {Burke}, {Boesgaard},
  {Chambers}, {Chun}, {Heasley}, {Hodapp}, {Hunt}, {Jedicke}, {Jewitt},
  {Kudritzki}, {Luppino}, {Maberry}, {Magnier}, {Monet}, {Onaka}, {Pickles},
  {Rhoads}, {Simon}, {Szalay}, {Szapudi}, {Tholen}, {Tonry}, {Waterson}, \&
  {Wick}}]{PSTARRS}
{Kaiser}, N., {Aussel}, H., {Burke}, B.~E., {Boesgaard}, H., {Chambers}, K.,
  {Chun}, M.~R., {Heasley}, J.~N., {Hodapp}, K.-W., {Hunt}, B., {Jedicke}, R.,
  {Jewitt}, D., {Kudritzki}, R., {Luppino}, G.~A., {Maberry}, M., {Magnier},
  E., {Monet}, D.~G., {Onaka}, P.~M., {Pickles}, A.~J., {Rhoads}, P.~H.~H.,
  {Simon}, T., {Szalay}, A., {Szapudi}, I., {Tholen}, D.~J., {Tonry}, J.~L.,
  {Waterson}, M., \& {Wick}, J. 2002, in Society of Photo-Optical
  Instrumentation Engineers (SPIE) Conference Series, Vol. 4836, Society of
  Photo-Optical Instrumentation Engineers (SPIE) Conference Series, ed.
  {J.~A.~Tyson \& S.~Wolff}, 154--164

\bibitem[{{Kasen} \& {Woosley}(2009)}]{Kasen09}
{Kasen}, D. \& {Woosley}, S.~E. 2009, \apj, 703, 2205

\bibitem[{{Kasen} {et~al.}(2011){Kasen}, {Woosley}, \& {Heger}}]{kasen11}
{Kasen}, D., {Woosley}, S.~E., \& {Heger}, A. 2011, \apj, 734, 102

\bibitem[{{Katz} {et~al.}(2012){Katz}, {Sapir}, \& {Waxman}}]{Katz11}
{Katz}, B., {Sapir}, N., \& {Waxman}, E. 2012, \apj, 747, 147

\bibitem[{{Klein} \& {Chevalier}(1978)}]{kl78}
{Klein}, R.~I. \& {Chevalier}, R.~A. 1978, \apjl, 223, L109

\bibitem[{{Kromer} \& {Sim}(2009)}]{KSim09}
{Kromer}, M. \& {Sim}, S.~A. 2009, \mnras, 398, 1809

\bibitem[{{Law} {et~al.}(2009){Law}, {Kulkarni}, {Dekany}, {Ofek}, {Quimby},
  {Nugent}, {Surace}, {Grillmair}, {Bloom}, {Kasliwal}, {Bildsten}, {Brown},
  {Cenko}, {Ciardi}, {Croner}, {Djorgovski}, {van Eyken}, {Filippenko}, {Fox},
  {Gal-Yam}, {Hale}, {Hamam}, {Helou}, {Henning}, {Howell}, {Jacobsen},
  {Laher}, {Mattingly}, {McKenna}, {Pickles}, {Poznanski}, {Rahmer}, {Rau},
  {Rosing}, {Shara}, {Smith}, {Starr}, {Sullivan}, {Velur}, {Walters}, \&
  {Zolkower}}]{PTF}
{Law}, N.~M., {Kulkarni}, S.~R., {Dekany}, R.~G., {Ofek}, E.~O., {Quimby},
  R.~M., {Nugent}, P.~E., {Surace}, J., {Grillmair}, C.~C., {Bloom}, J.~S.,
  {Kasliwal}, M.~M., {Bildsten}, L., {Brown}, T., {Cenko}, S.~B., {Ciardi}, D.,
  {Croner}, E., {Djorgovski}, S.~G., {van Eyken}, J., {Filippenko}, A.~V.,
  {Fox}, D.~B., {Gal-Yam}, A., {Hale}, D., {Hamam}, N., {Helou}, G., {Henning},
  J., {Howell}, D.~A., {Jacobsen}, J., {Laher}, R., {Mattingly}, S., {McKenna},
  D., {Pickles}, A., {Poznanski}, D., {Rahmer}, G., {Rau}, A., {Rosing}, W.,
  {Shara}, M., {Smith}, R., {Starr}, D., {Sullivan}, M., {Velur}, V.,
  {Walters}, R., \& {Zolkower}, J. 2009, \pasp, 121, 1395

\bibitem[{{Magee} {et~al.}(1995){Magee}, {Abdallah}, {Clark}, {Cohen},
  {Collins}, {Csanak}, {Fontes}, {Gauger}, {Keady}, {Kilcrease}, \&
  {Merts}}]{oplib}
{Magee}, N.~H., {Abdallah}, Jr., J., {Clark}, R.~E.~H., {Cohen}, J.~S.,
  {Collins}, L.~A., {Csanak}, G., {Fontes}, C.~J., {Gauger}, A., {Keady},
  J.~J., {Kilcrease}, D.~P., \& {Merts}, A.~L. 1995, in Astronomical Society of
  the Pacific Conference Series, Vol.~78, Astrophysical Applications of
  Powerful New Databases, ed. {S.~J.~Adelman \& W.~L.~Wiese}, 51

\bibitem[{{Matzner} \& {McKee}(1999)}]{mm99}
{Matzner}, C.~D. \& {McKee}, C.~F. 1999, \apj, 510, 379

\bibitem[{{Mazzali} \& {Lucy}(1993)}]{ML93}
{Mazzali}, P.~A. \& {Lucy}, L.~B. 1993, \aap, 279, 447

\bibitem[{{Moriya} {et~al.}(2010){Moriya}, {Yoshida}, {Tominaga}, {Blinnikov},
  {Maeda}, {Tanaka}, \& {Nomoto}}]{moriya10}
{Moriya}, T., {Yoshida}, N., {Tominaga}, N., {Blinnikov}, S.~I., {Maeda}, K.,
  {Tanaka}, M., \& {Nomoto}, K. 2010, in American Institute of Physics
  Conference Series, Vol. 1294, American Institute of Physics Conference
  Series, ed. {D.~J.~Whalen, V.~Bromm, \& N.~Yoshida}, 268--269

\bibitem[{{Nakar} \& {Sari}(2010)}]{ns10}
{Nakar}, E. \& {Sari}, R. 2010, \apj, 725, 904

\bibitem[{{Ofek} {et~al.}(2010){Ofek}, {Rabinak}, {Neill}, {Arcavi}, {Cenko},
  {Waxman}, {Kulkarni}, {Gal-Yam}, {Nugent}, {Bildsten}, {Bloom}, {Filippenko},
  {Forster}, {Howell}, {Jacobsen}, {Kasliwal}, {Law}, {Martin}, {Poznanski},
  {Quimby}, {Shen}, {Sullivan}, {Dekany}, {Rahmer}, {Hale}, {Smith},
  {Zolkower}, {Velur}, {Walters}, {Henning}, {Bui}, \& {McKenna}}]{ofek10}
{Ofek}, E.~O., {Rabinak}, I., {Neill}, J.~D., {Arcavi}, I., {Cenko}, S.~B.,
  {Waxman}, E., {Kulkarni}, S.~R., {Gal-Yam}, A., {Nugent}, P.~E., {Bildsten},
  L., {Bloom}, J.~S., {Filippenko}, A.~V., {Forster}, K., {Howell}, D.~A.,
  {Jacobsen}, J., {Kasliwal}, M.~M., {Law}, N., {Martin}, C., {Poznanski}, D.,
  {Quimby}, R.~M., {Shen}, K.~J., {Sullivan}, M., {Dekany}, R., {Rahmer}, G.,
  {Hale}, D., {Smith}, R., {Zolkower}, J., {Velur}, V., {Walters}, R.,
  {Henning}, J., {Bui}, K., \& {McKenna}, D. 2010, \apj, 724, 1396

\bibitem[{{Piro} {et~al.}(2010){Piro}, {Chang}, \& {Weinberg}}]{Piro10}
{Piro}, A.~L., {Chang}, P., \& {Weinberg}, N.~N. 2010, \apj, 708, 598

\bibitem[{{Pomraning}(1991)}]{Pom91}
{Pomraning}, G.~C. 1991, {Linear Kinetic Theory and Particle Transport in
  Stochastic Mixtures}

\bibitem[{{Roming} {et~al.}(2012){Roming}, {Pritchard}, {Prieto}, {Kochanek},
  {Fryer}, {Davidson}, {Humphreys}, {Bayless}, {Beacom}, {Brown}, {Holland},
  {Immler}, {Kuin}, {Oates}, {Pogge}, {Pojmanski}, {Stoll}, {Shappee},
  {Stanek}, \& {Szczygiel}}]{Rom12}
{Roming}, P.~W.~A., {Pritchard}, T.~A., {Prieto}, J.~L., {Kochanek}, C.~S.,
  {Fryer}, C.~L., {Davidson}, K., {Humphreys}, R.~M., {Bayless}, A.~J.,
  {Beacom}, J.~F., {Brown}, P.~J., {Holland}, S.~T., {Immler}, S., {Kuin},
  N.~P.~M., {Oates}, S.~R., {Pogge}, R.~W., {Pojmanski}, G., {Stoll}, R.,
  {Shappee}, B.~J., {Stanek}, K.~Z., \& {Szczygiel}, D.~M. 2012, \apj, 751, 92

\bibitem[{{Schawinski} {et~al.}(2008){Schawinski}, {Justham}, {Wolf},
  {Podsiadlowski}, {Sullivan}, {Steenbrugge}, {Bell}, {R{\"o}ser}, {Walker},
  {Astier}, {Balam}, {Balland}, {Carlberg}, {Conley}, {Fouchez}, {Guy},
  {Hardin}, {Hook}, {Howell}, {Pain}, {Perrett}, {Pritchet}, {Regnault}, \&
  {Yi}}]{sch08}
{Schawinski}, K., {Justham}, S., {Wolf}, C., {Podsiadlowski}, P., {Sullivan},
  M., {Steenbrugge}, K.~C., {Bell}, T., {R{\"o}ser}, H.-J., {Walker}, E.~S.,
  {Astier}, P., {Balam}, D., {Balland}, C., {Carlberg}, R., {Conley}, A.,
  {Fouchez}, D., {Guy}, J., {Hardin}, D., {Hook}, I., {Howell}, D.~A., {Pain},
  R., {Perrett}, K., {Pritchet}, C., {Regnault}, N., \& {Yi}, S.~K. 2008,
  Science, 321, 223

\bibitem[{{Smith}(2003)}]{Smith03}
{Smith}, C.~C. 2003, Journal of Quantitative Spectroscopy and Radiative
  Transfer, 81, 451

\bibitem[{{Smith} \& {McCray}(2007)}]{nsmith07}
{Smith}, N. \& {McCray}, R. 2007, \apjl, 671, L17

\bibitem[{{Soderberg} {et~al.}(2008){Soderberg}, {Berger}, {Page}, {Schady},
  {Parrent}, {Pooley}, {Wang}, {Ofek}, {Cucchiara}, {Rau}, {Waxman}, {Simon},
  {Bock}, {Milne}, {Page}, {Barentine}, {Barthelmy}, {Beardmore}, {Bietenholz},
  {Brown}, {Burrows}, {Burrows}, {Byrngelson}, {Cenko}, {Chandra}, {Cummings},
  {Fox}, {Gal-Yam}, {Gehrels}, {Immler}, {Kasliwal}, {Kong}, {Krimm},
  {Kulkarni}, {Maccarone}, {M{\'e}sz{\'a}ros}, {Nakar}, {O'Brien}, {Overzier},
  {de Pasquale}, {Racusin}, {Rea}, \& {York}}]{sod08}
{Soderberg}, A.~M., {Berger}, E., {Page}, K.~L., {Schady}, P., {Parrent}, J.,
  {Pooley}, D., {Wang}, X.-Y., {Ofek}, E.~O., {Cucchiara}, A., {Rau}, A.,
  {Waxman}, E., {Simon}, J.~D., {Bock}, D.~C.-J., {Milne}, P.~A., {Page},
  M.~J., {Barentine}, J.~C., {Barthelmy}, S.~D., {Beardmore}, A.~P.,
  {Bietenholz}, M.~F., {Brown}, P., {Burrows}, A., {Burrows}, D.~N.,
  {Byrngelson}, G., {Cenko}, S.~B., {Chandra}, P., {Cummings}, J.~R., {Fox},
  D.~B., {Gal-Yam}, A., {Gehrels}, N., {Immler}, S., {Kasliwal}, M., {Kong},
  A.~K.~H., {Krimm}, H.~A., {Kulkarni}, S.~R., {Maccarone}, T.~J.,
  {M{\'e}sz{\'a}ros}, P., {Nakar}, E., {O'Brien}, P.~T., {Overzier}, R.~A., {de
  Pasquale}, M., {Racusin}, J., {Rea}, N., \& {York}, D.~G. 2008, \nat, 453,
  469

\bibitem[{{Tolstov}(2010)}]{Tolstov10}
{Tolstov}, A.~G. 2010, Astronomy Letters, 36, 109

\bibitem[{{Tominaga} {et~al.}(2009){Tominaga}, {Blinnikov}, {Baklanov},
  {Morokuma}, {Nomoto}, \& {Suzuki}}]{Tomin09}
{Tominaga}, N., {Blinnikov}, S., {Baklanov}, P., {Morokuma}, T., {Nomoto}, K.,
  \& {Suzuki}, T. 2009, \apjl, 705, L10

\bibitem[{{van Marle} {et~al.}(2010){van Marle}, {Smith}, {Owocki}, \& {van
  Veelen}}]{vmarle10}
{van Marle}, A.~J., {Smith}, N., {Owocki}, S.~P., \& {van Veelen}, B. 2010,
  \mnras, 407, 2305

\bibitem[{{Waxman} {et~al.}(2007){Waxman}, {M{\'e}sz{\'a}ros}, \&
  {Campana}}]{wax07}
{Waxman}, E., {M{\'e}sz{\'a}ros}, P., \& {Campana}, S. 2007, \apj, 667, 351

\bibitem[{{Weaver} {et~al.}(1978){Weaver}, {Zimmerman}, \&
  {Woosley}}]{Weaver1978}
{Weaver}, T.~A., {Zimmerman}, G.~B., \& {Woosley}, S.~E. 1978, \apj, 225, 1021

\bibitem[{{Whalen} {et~al.}(2012{\natexlab{a}}){Whalen}, {Frey}, {Even},
  {Fryer}, {Heger}, {Holz}, {Hungerford}, {Lovekin}, \& {Stiavelli}}]{wet12a}
{Whalen}, D.~J., {Frey}, L., {Even}, W., {Fryer}, C., {Heger}, A., {Holz}, D.,
  {Hungerford}, A., {Lovekin}, C., \& {Stiavelli}, M. 2012{\natexlab{a}}, apJ,
  submitted (arXiv:1211.4979)

\bibitem[{{Whalen} {et~al.}(2013){Whalen}, {Fryer}, {Holz}, {Heger}, {Woosley},
  {Stiavelli}, {Even}, \& {Frey}}]{wet12c}
{Whalen}, D.~J., {Fryer}, C.~L., {Holz}, D.~E., {Heger}, A., {Woosley}, S.~E.,
  {Stiavelli}, M., {Even}, W., \& {Frey}, L.~H. 2013, \apjl, 762, L6

\bibitem[{{Whalen} {et~al.}(2012{\natexlab{b}}){Whalen}, {Joggerst}, {Fryer},
  {Stiavelli}, {Heger}, \& {Holz}}]{wet12b}
{Whalen}, D.~J., {Joggerst}, C., {Fryer}, C., {Stiavelli}, M., {Heger}, A., \&
  {Holz}, D. 2012{\natexlab{b}}, arXiv:1209.5459

\bibitem[{{Woosley} {et~al.}(2002){Woosley}, {Heger}, \&
  {Weaver}}]{Woosley2002}
{Woosley}, S.~E., {Heger}, A., \& {Weaver}, T.~A. 2002, Reviews of Modern
  Physics, 74, 1015

\bibitem[{{Young} \& {Arnett}(2005)}]{ya05}
{Young}, P.~A. \& {Arnett}, D. 2005, \apj, 618, 908

\bibitem[{{Young} {et~al.}(2006){Young}, {Fryer}, {Hungerford}, {Arnett},
  {Rockefeller}, {Timmes}, {Voit}, {Meakin}, \& {Eriksen}}]{y06}
{Young}, P.~A., {Fryer}, C.~L., {Hungerford}, A., {Arnett}, D., {Rockefeller},
  G., {Timmes}, F.~X., {Voit}, B., {Meakin}, C., \& {Eriksen}, K.~A. 2006,
  \apj, 640, 891

\bibitem[{{Zeeb} \& {Fontes}(2006)}]{tops06}
{Zeeb}, C.~N. \& {Fontes}, C.~J. 2006, los Alamos National Laboratory Report
  No. LA-UR-06-0882 (unpublished)

\bibitem[{{Zhang} {et~al.}(2008){Zhang}, {Woosley}, \& {Heger}}]{zwh08}
{Zhang}, W., {Woosley}, S.~E., \& {Heger}, A. 2008, \apj, 679, 639

\end{thebibliography}

\end{document}